\documentclass[conference]{IEEEtran}
\usepackage[utf8]{inputenc}
\usepackage{xcolor}
\usepackage{array}
\usepackage{tabularx}
\usepackage{booktabs}
\usepackage{multirow}
\usepackage{multicol}
\usepackage{rotating}
\usepackage{hyperref}
\usepackage[capitalise]{cleveref}
\usepackage{cite}
\usepackage{tikz}
\usepackage{adjustbox}
\usepackage{makecell}
\usepackage{pdflscape}
\usepackage{caption}
\usepackage{subcaption}
\usepackage{threeparttable}

\usepackage[all=normal,floats=tight,paragraphs=tight]{savetrees}

\title{SoK: Hardware-supported Trusted Execution Environments}
 \author{
     \IEEEauthorblockN{Moritz Schneider}
     \IEEEauthorblockA{ETH Zurich}
     \and
     \IEEEauthorblockN{Ramya Jayaram Masti}
     \IEEEauthorblockA{Intel Cooperation}
     \and
     \IEEEauthorblockN{Shweta Shinde}
     \IEEEauthorblockA{ETH Zurich}
     \and
     \IEEEauthorblockN{Srdjan Capkun}
     \IEEEauthorblockA{ETH Zurich}
     \and
     \IEEEauthorblockN{Ronald Perez}
     \IEEEauthorblockA{Intel Cooperation}
 }
\date{August 2021}

\usepackage{xspace}
\newcommand{\enclave}{enclave\xspace}
\newcommand{\Enclave}{Enclave\xspace}

\newcommand{\advTEE}{{$A_{tee}$\xspace}}
\newcommand{\advApp}{{$A_{app}$\xspace}}
\newcommand{\advBoot}{{$A_{boot}$\xspace}}
\newcommand{\advSSW}{{$A_{ssw}$\xspace}}
\newcommand{\advPer}{{$A_{per}$\xspace}}

\newcommand{\advCPU}{{$A_{inv}$\xspace}}

\newcommand{\advBus}{{$A_{bus}$\xspace}}

\definecolor{blue}{HTML}{4477AA}
\definecolor{cyan}{HTML}{66CCEE}
\definecolor{green}{HTML}{228833}
\definecolor{yellow}{HTML}{CCBB44}
\definecolor{red}{HTML}{EE6677}
\definecolor{purple}{HTML}{AA3377}
\definecolor{grey}{HTML}{BBBBBB}

\definecolor{dark grey}{HTML}{555555}

\definecolor{light blue}{HTML}{77AADD}
\definecolor{light cyan}{HTML}{99DDFF}
\definecolor{mint}{HTML}{44bb00}
\definecolor{pear}{HTML}{BBCC33}
\definecolor{olive}{HTML}{AAAA00}
\definecolor{light yellow}{HTML}{EEDD88}
\definecolor{orange}{HTML}{EE8866}
\definecolor{pink}{HTML}{FFAABB}
\definecolor{light grey}{HTML}{DDDDDD}

\definecolor{pale grey}{HTML}{DDDDDD}
\definecolor{pale green}{HTML}{CCDDAA}
\definecolor{pale red}{HTML}{FFCCCC}
\definecolor{pale cyan}{HTML}{CCEEFF}
\definecolor{pale blue}{HTML}{BBCCEE}
\definecolor{pale grey}{HTML}{DDDDDD}

\definecolor{vibrant red}{HTML}{CC3311}

\graphicspath{{figures/}}

\newif\ifcomments{}
\commentsfalse{}

\ifcomments{}
\newcommand{\ramya}[1]{\textbf{\emph{ #1 \colorbox{green}{[Ramya]}}}}
\newcommand{\shweta}[1]{\textbf{\emph{ #1 \colorbox{cyan}{[Shweta]}}}}
\newcommand{\moritz}[1]{\textbf{\emph{ #1 \colorbox{yellow}{[Moritz]}}}}
\newcommand{\srdjan}[1]{\textbf{\emph{ #1 \colorbox{blue}{[Srdjan]}}}}
\newcommand{\todo}[1]{\textcolor{vibrant red}{TODO: \@#1}}
\newcommand{\todoref}{\textcolor{vibrant red}{[ref]}}
\newcommand{\citneed}{[\textcolor{vibrant red}{cit}] }

\else
\newcommand{\ramya}[1]{}
\newcommand{\shweta}[1]{}
\newcommand{\moritz}[1]{}
\newcommand{\srdjan}[1]{}
\newcommand{\todo}[1]{}
\newcommand{\todoref}[1]{}
\newcommand{\citneed}[1]{}

\fi

\usetikzlibrary{tikzmark,positioning,shapes,calc}
\tikzstyle{every node}=[font=\small]

\newcommand*{\circl}[1]{\begin{tikzpicture}[scale=0.11]%
    \draw (0,0) circle (1);
    \fill[fill=black!100] (0,0) -- (90:1) arc (90:90-#1*3.6:1) -- cycle;
    \end{tikzpicture}}

\newcommand{\yes}{\circl{100}}
\newcommand{\no}{\circl{0}}

\begin{document}

\maketitle
\pagestyle{plain}
\begin{abstract}
The growing complexity of modern computing platforms and the need for strong isolation protections among their software components has led to the increased adoption of Trusted Execution Environments (TEEs). While several commercial and academic TEE architectures have emerged in recent times, they remain hard to compare and contrast. More generally, existing TEEs have not been subject to a holistic systematization to understand the available design alternatives for various aspects of TEE design and their corresponding pros-and-cons. %

Therefore, in this work, we analyze the design of existing TEEs and systematize the mechanisms that TEEs implement to achieve their security goals, namely, verifiable launch, run-time isolation, trusted IO and secure storage. More specifically, we analyze the typical architectural building blocks underlying TEE solutions, design alternatives for each of these components and the trade-offs that they entail. We focus on hardware-assisted TEEs and cover a wide range of TEE proposals from academia and the industry. Our analysis shows that although TEEs are diverse in terms of their goals, usage models and instruction set architectures, they all share many common building blocks in terms of their design.
\end{abstract}
\section{Introduction}

Today's computing platforms are diverse in terms of their architectures, software provisioning models, and the types of applications they support. They include large-scale servers used in cloud computing, smartphones used in mobile networks, and smart home devices. Over time, these devices have evolved to store and process security-sensitive data to enable novel applications in various domains such as the healthcare and financial industry. Therefore, modern computing platforms must implement mechanisms to protect such security-sensitive data against unauthorized access and modification. 

Data confidentiality and integrity solutions in today's computing systems have to not only account for attacks over the network, but also those that originate from (a subset of) software and hardware components on the same platform or from an adversary with physical access. This is because these systems typically host multiple, mutually-untrusted software components. Examples of such software deployment models include code/data from different tenants that share the same cloud platform and code/data belonging to users and network providers that share the same smartphone. Therefore, platforms today have to isolate sensitive data from potentially co-resident software and hardware adversaries. In fact, this requirement equally extends to any computations that involve such sensitive data and, more generally, any security-critical computations.

An increasingly popular solution to the problem of protecting sensitive computations and data against co-located attackers is a \emph{Trusted Execution Environment (TEE)}. While existing  TEEs often vary in terms of their exact security goals, most of them aim to provide (a subset of) four high level security protections, namely, \emph{(i)} verifiable launch of the execution environment for the sensitive code and data so that a remote entity can ensure that it was set up correctly, \emph{(ii)} run-time isolation to protect the confidentiality and integrity of sensitive code and data, \emph{(iii)} trusted IO to enable secure access to peripherals and accelerators, and finally, \emph{(iv)} secure storage for TEE data that must be stored persistently and made available only to authorized entities at a later point in time.

\begin{figure}
    \centering
    \includegraphics[width=1\linewidth]{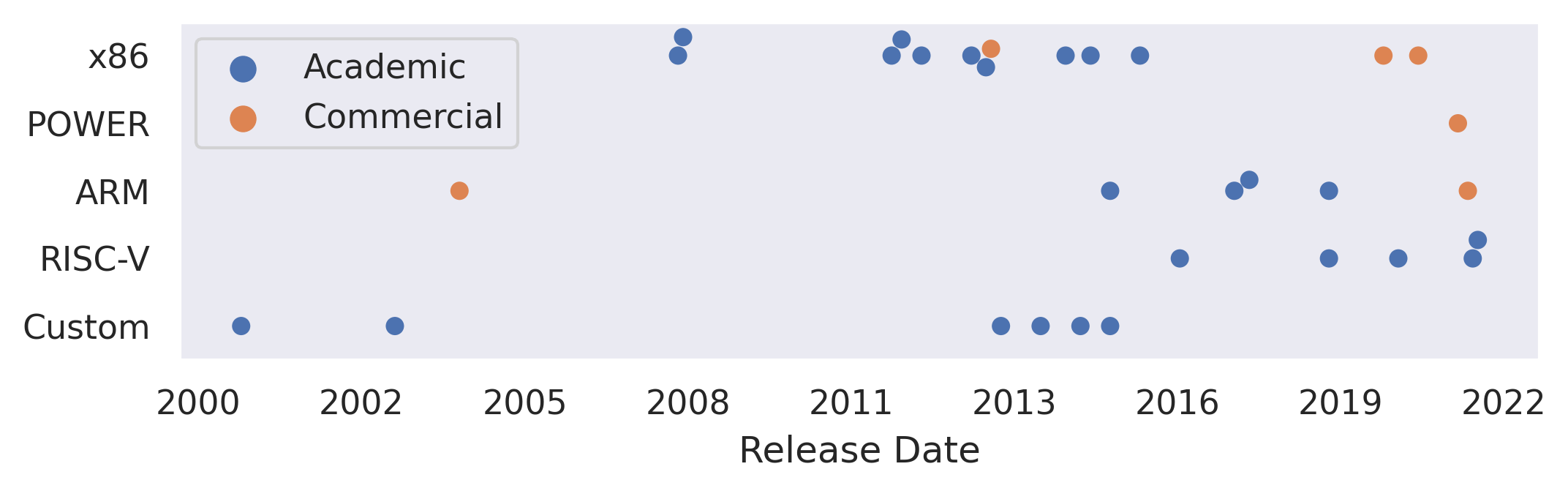}
    \caption{The release dates of the surveyed TEEs according to their instruction set architecture.}
    \label{fig:date}
\end{figure}

Today, a variety of academic and commercial TEE designs and solutions exist. They are diverse in terms of their underlying security assumptions, their target instruction set architectures and the usage models they support (\cref{fig:date}). 
In this work, we analyze the design of existing TEEs and systematize common as well as unique design decisions of TEEs. Even though many TEE designs use different names and descriptions of their underlying mechanisms, the resulting mechanisms are often very similar. To reason about the similarities between these designs, we group the underlying mechanisms into classes of mechanisms and highlight exceptional cases.

We begin our study of TEE solutions by identifying and classifying the common set of adversaries that most TEEs consider, based on which platform components they control in software and hardware. %
We focus on hardware-assisted TEEs and cover how they implement verifable launch, run-time CPU and memory isolation, trusted IO and secure storage. Analyzing which of these features different TEEs support, how they implement them, and which attackers they consider allows us to compare a wide range of TEEs spanning different usages including mobile and cloud computing. It also enables us to be inclusive and cover a vast majority of existing academic and commercial TEEs.

The first TEE security goal we study is verifiable launch which refers to providing proof regarding the correctness of the initial state of the TEE. This is typically achieved by first establishing a Root of Trust for Measurement (RTM), then leveraging it to \emph{measure} the state of the code/data within the TEE, and finally, making this available for verification through a process called \emph{attestation}.  Standard measurement and attestation processes have long been established by the Trusted Computing Group using a Trusted Platform Module (TPM)\cite{tpm}. To a large extent, today's attestation solutions involve similar mechanisms and protocols but have evolved over time to rely on different architectural components. For example, in some TEEs, attestation keys and measurements are held in on-chip components\cite{costan2016intel,kaplan2020amd,costan2016sanctum}, but others rely on the off-chip TPM\cite{mccune2008flicker,mccune2008low}. Key hierarchies involved in modern attestation protocols are also different from standard TPM-based protocols and, in some cases, involve symmetric keys for within-platform attestation for performance reasons. %

We then focus on how TEEs implement run-time isolation to protect the confidentiality and integrity of sensitive computations and data. We introduce a taxonomy of isolation strategies that classifies them according to two dimensions: resource partitioning and isolation enforcement. We then use these dimensions to reason about the techniques used by individual TEE designs. As each strategy has unique  (dis-)advantages for every resource, we discuss their suitability for isolating CPU and memory against different adversaries. Then, we summarize if and how different TEE solutions adopt these strategies by describing the various architectural  components they use for CPU and memory isolation. We show that despite their diversity, most modern TEE designs use a single common strategy for CPU protection. In contrast, the strategies for memory protection are diverse, with some TEEs employing different strategies against different adversaries.

Trusted IO solutions for use with TEEs have evolved over time from supporting user IO to more diverse accelerators. Most trusted IO solutions involve two main components: a trusted path to the device and a trusted device architecture to protect security sensitive data just like the CPU. We identify two common types of trusted paths that TEEs can implement, namely logical and cryptographic. Then, we describe different ways to implement these trusted paths, their suitability for use in different scenarios and the architectural support they require in each case. Finally, we apply the same taxonomy of isolation strategies that we used in the context of CPU and memory isolation to understand diverse trusted device architectures.

Secure storage in the context of TEEs involves ensuring that any sensitive data that is persistent is only available to authorized entities. This concept is often referred to as \emph{sealing}. Similar to measurement and attestation solutions, early sealing mechanisms for TEEs typically rely on concepts pioneered by the TPM\cite{tpm}. We observe that sealing processes have adapted over time to cater to new requirements (e.g., migration, anti-rollback) as well as rely only on on-chip elements. Our study reveals that only about a third of existing TEE solutions discuss sealing support with similar implementation approaches.

Overall, this paper makes the following contributions towards an improved understanding of TEE architectures.

\begin{itemize}
    \item This paper describes an adversary model including software and physical attackers and their capabilities. We believe that this taxonomy of adversaries is useful both during the design of TEEs in terms of choosing between different security mechanisms as well as for analyzing their security. This stems from the fact that the choice of security mechanisms in TEE solutions often differ not only by the resource being protected but also by the type of adversary being thwarted.
    \item To the best of our knowledge, this is the first effort to identify and classify the design decisions made by TEEs to achieve four fundamental security goals, namely verifiable launch, run-time isolation, trusted IO, and secure storage. We believe that this systematization can be used as a basis for designing new TEE architectures based on different design constraints and security models.
    \item Based on the surveyed TEE architectures, we conclude that the design space of the four fundamental security goals to be relatively small. New proposals usually re-use many design choices and only propose minor modifications to a single component. \todo{More major findings}
\end{itemize}

\section{Scope}

Numerous new TEE architectures have been proposed in recent years. Oddly, despite the abundant research on TEEs and the growing number of commercially available TEE solutions, there is no single, widely-accepted definition for a TEE. In this paper, we do not attempt to find such a general definition of a TEE. Instead, we survey a wide variety of existing approaches and systematize them according to their architectural support of four security properties common in all of them: (i) verifiable launch, (ii) run-time isolation, (iii) trusted IO, and (iv) secure storage. With all the differences in existing TEE designs, we aim to find underlying design decisions that connect all these seemingly different approaches. Since the performance of a TEE is mainly determined by the processor design and not the TEE specifics, we do not investigate performance differences between the surveyed TEEs.

The selection of designs to survey is critical to this paper and it is based on the following criteria:
\begin{itemize}
    \item We focus on hardware-assisted TEEs and do not investigate arguably more complex software approaches, e.g., relying on trusted hypervisors.
    \item We consider TEEs from many different processor architectures. We selected designs covering four major processor architectures: x86, ARM, POWER, and RISC-V. At the same time we also analyzed proposals based on more niche instruction set architectures such as SPARC and OpenRISC.
    \item While some proposals are sometimes not regarded to be TEEs (e.g., ARM TrustZone) we still attempt to include as many designs that may fit in the envelope of a TEE even if the proposal itself does not use the term ``TEE''.
    \item We deliberately exclude proposals based on co-processors such as Google Titan and Apple's Secure Enclaves, or proposals based on hardware security modules (HSM). We want to focus on approaches that run on general-purpose processors and are closely intertwined with other untrusted applications running on the same processor. %
\end{itemize}

\noindent \textbf{Terminology:} In academic literature and the industry, numerous names have been used for the various components of a TEE. 
For example, Intel Software Guard Extensions (SGX) refers to TEE instances as \emph{enclaves}~\cite{mckeen2013innovative}, Trust Domain Extensions (TDX) uses \emph{Trust Domains}~\cite{tdxbasespec}, AMD Secure Encrypted VMs-Secure Nested Paging (SEV-SNP)\cite{kaplan2020amd} uses \emph{Secure Encrypted VMs}, ARM Confidential Compute Architecture (CCA) uses \emph{Realms} for their VM-isolation solution and \emph{trustlets} for their application isolation feature\cite{armrealms}. In this paper, we use \emph{\enclave{}} to refer to such a TEE instance and trusted computing base (\emph{TCB}) to describe all underlying trusted components. We use \emph{TEE} to refer to the entire architecture that enables the creation of \enclave{s}. 
\section{System and Adversary Models}
This section discusses the generic platform underlying most TEE designs and the types of adversaries they consider.

\subsection{System Model}
\begin{figure}[tbp]
    \centering
    \includegraphics[width=0.95\linewidth]{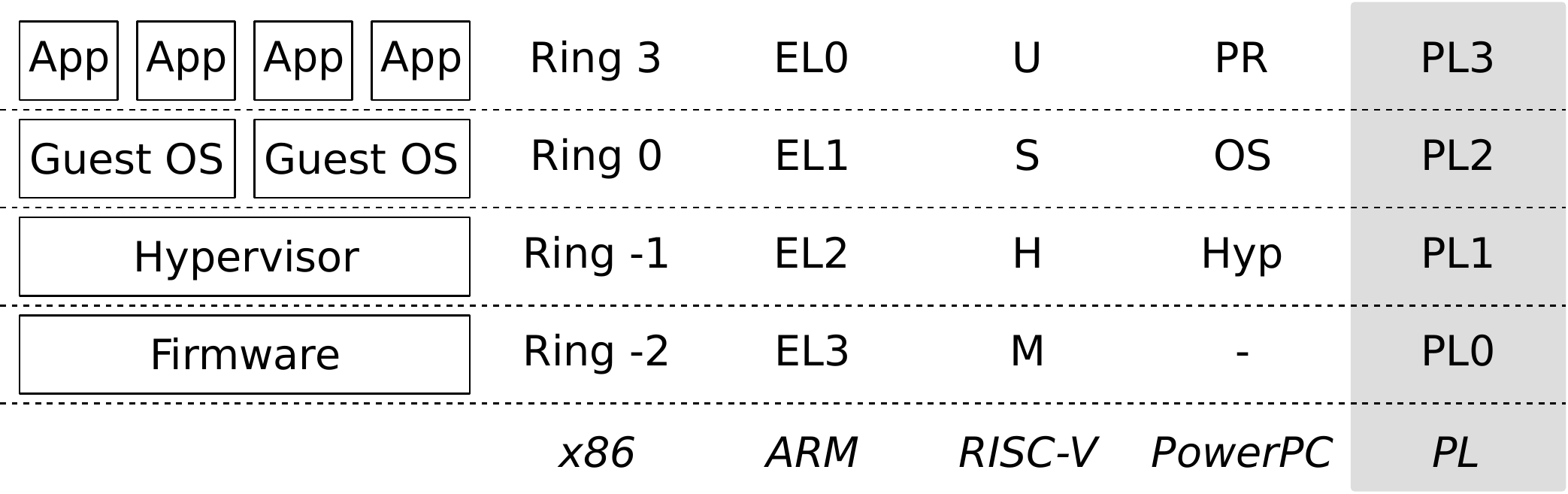}
    \caption{Summary of privilege levels in modern processors: Most CPUs support at least four privilege levels, one each for userspace applications, an OS, a hypervisor. In this paper, we will use PL0-PL3 shown in grey to denote them.}
    \label{fig:privlevel}
\end{figure}

Most TEE solutions target a modern general-purpose computing platform that consists of a System-on-Chip (SoC) with off-chip memory and, optionally, off-chip peripherals, as shown in \cref{fig:adversaries}. The SoC itself contains one or more cores that potentially share a cache (or a part thereof) and fabric that connects them to the memory controller. SoCs  also include an IO complex that is used to connect both on-chip and off-chip peripherals to the SoC fabric and caches. The typical software stack of the system includes an operating system (OS) and multiple userspace applications. If virtualized, a hypervisor runs underneath one or more Virtual Machines (VMs), each with their own (guest) OS and userspace applications. These software components run at different privilege levels on most modern CPUs (see \cref{fig:privlevel}). The hardware and software components that are used to achieve the security protections of a TEE is called the Trusted Computing Base (TCB). %

\subsection{Adversary Model}
\label{sec:adversaries}
\begin{figure}[tbp]
    \centering
    \includegraphics[width=0.8\linewidth]{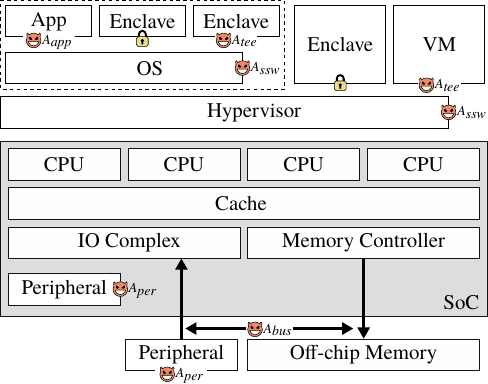}
    \caption{The software and hardware system model with our adversary model. The boot adversary (\advBoot) and the invasive adversary (\advCPU) are omitted.}
    \label{fig:adversaries}
\end{figure}

TEEs aim to provide a variety of security protections against a wide range of adversaries that are co-resident on the same platform. 
These include untrusted co-resident software (e.g., code in other \enclave{}s, system management software such as an OS), untrusted platform hardware (e.g., IO  peripherals), a hardware attacker who has physical access to the platform  (e.g., bus interposers) or a combination thereof. We discuss these adversaries and the types of attacks they can launch with respect to a \emph{victim \enclave} (see \cref{fig:adversaries}). 

\subsubsection*{Co-located \enclave{} adversary} 
This adversary is relevant when the platform supports more than one \enclave{}, either concurrently or over time. This adversary is capable of launching one or more \enclave{}s with code/data of its choice. Such a situation commonly emerges, for example, in multi-tenant cloud platforms where multiple users can launch \enclave{}s on shared hardware. It also occurs in mobile ecosystems where multiple service providers provision code (and data) to run within enclaves. For brevity, we use \advTEE{} to refer to such an attacker as well as any code/resources that it controls.

\subsubsection*{Unprivileged software adversary}
This adversary can launch any \emph{unprivileged} software on the same shared hardware as the victim \enclave{}. Such an attacker can often run code with the same privilege level as the victim's \enclave but not higher. Examples of such adversaries include cloud users that control guest VMs in cloud environments that run alongside the victim \enclave, and mobile phone users that can launch apps to run concurrently with mobile phone \enclave{}s. We use \advApp~to refer to such an attacker. 

\subsubsection*{System software adversary}
This refers to an adversary that controls the system management software, such as an OS that manages the platform's resources. This adversary has all the capabilities of \advTEE~and \advApp. Additionally, it controls system resources such as memory and scheduling. Hence, it is more powerful than the above adversaries. Examples of such adversaries include the untrusted hypervisor in cloud settings and the OS running on a mobile phone. This adversary is referred to as \advSSW{}.

\subsubsection*{Startup adversary}
This refers to an adversary that controls the system boot of the platform that hosts the TEE. Such an attacker controls system configuration such as memory and IO fabric parameters that could undermine the entire TEE design if misconfigured. Examples of such an attacker include an untrusted BIOS. This adversary is represented as \advBoot{}.

\subsubsection*{Peripheral adversary}
Modern platforms could include multiple peripherals that are not in the TCB of the victim \enclave{}. These peripherals could be within the SoC or the connected over to it over an external bus. They are often assumed to be untrusted, especially if they include firmware that could be potentially exploited  remotely. An adversary controlling such a peripheral can launch nefarious IO transactions to try to access or modify memory and other resources belonging to the victim \enclave. We use \advPer{} to denote such an attacker.

\subsubsection*{Fabric adversary} 
Another adversary considered by TEE architectures has the ability to introduce special hardware such as fabric interposers to launch man-in-the-middle attacks\cite{lee2020off}. The fabric adversary can also directly access data-at-rest such as the disk or external memory. However, this adversary cannot breach the SoC package: everything within the package remains out-of-scope. 
In the rest of the paper, we  use \advBus{} to represent this adversary.

\subsubsection*{Invasive adversary}
This adversary can launch invasive attacks such as de-layering the physical chip, manipulating clock signals and voltage rails to cause faults, etc., to extract secrets or force a  different execution path than the intended one. For the sake of completeness, we include this adversary (\advCPU) in our list but note that no TEE design currently defends against such an attacker. So, we do not discuss this attacker any further in this paper.

\subsection{A Note on Side-Channel Attacks}
Numerous physical~\cite{krachenfels2021laser,kocher1999differential} and micro-architectural \cite{liu2015last,brasser2017software,li2021cipherleaks} side-channel attacks have been explored and shown to be feasible on modern systems both in the context of TEEs\cite{brasser2017software} as well more broadly\cite{liu2015last}. Such attacks can be launched by any of the above adversaries with varying success. Given that side-channels are often a result of shared computation resources and not often specific to \enclave{}s, TEEs often rely on generic countermeasures to mitigate their impact. For example, generic software defense approaches such as eliminating secret-dependent memory accesses\cite{ahmad2019obfuscuro} or secret dependent branching\cite{rane2015raccoon} equally apply for use with \enclave{}s. Today, most (commercial) TEE proposals explicitly exclude side-channel attacks from their attacker model and recommend using existing countermeasures to protect against them. Even though side-channel attacks are not the focus of this paper, we will still briefly mention the impact of specific design decisions on side-channels where it is appropriate.

\section{Verifiable Launch}
\label{sec:attestation}

A critical first step that precedes the actual execution of an \enclave{} is a secure setup process that ensures that the \enclave{}'s execution environment is configured correctly and that its initial state is as expected. The prevalent verifiable boot process in TEEs is \enclave{} measurement and attestation. Intuitively, a measurement of an \enclave{}, and more generally any software, is a fingerprint of its initial state, typically constructed by a series of one or more cryptographic hashes over the initial memory of the \enclave{}. The measurement process itself must be trustworthy; it begins at the Root-of-Trust for Measurement (RTM) and is implemented as a chain-of-trust through the \enclave{}'s TCB, which finally measures the \enclave{} itself. The measurements are used later as part of a digitally signed report sent to a verifier through a process referred to as attestation. Attestation could provide additional information about the \enclave{}'s security properties (e.g., the authenticity of the platform hosting the \enclave{}, details about its TCB itself). This section discusses architectural support for different types of RTMs, typical measurement processes, and attestation schemes of TEEs. 

\begin{figure}
    \centering
    \begin{subfigure}[*]{0.9\linewidth}
        \centering
        \includegraphics[width=0.9\linewidth]{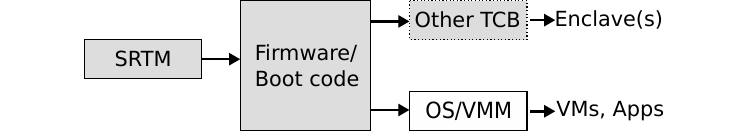}
        \caption{Static Root of Trust for Measurement (SRTM)}
        \label{fig:rtm:srtm}
    \end{subfigure}
    \par\bigskip
    \begin{subfigure}[*]{0.9\linewidth}
        \centering
        \includegraphics[width=0.9\linewidth]{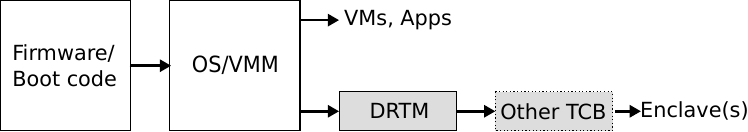}
        \caption{Dynamic Root of Trust for Measurement (DRTM)}
        \label{fig:rtm:drtm}
    \end{subfigure}
    \caption{Types of Root of Trust for Measurement (RTM) used by modern TEEs: The RTM may directly measure the \enclave or optionally measure and launch one or more TCB components that eventually measure the \enclave. The chain-of-trust for measurement (shown in grey) consists of the RTM and any intermediate TCB components that are eventually responsible for the \enclave{'s} final measurement.}
    \label{fig:rtm}
\end{figure}

\subsection{Root of Trust (RTM)}
\label{sec:attest:rtm}
Central to a verifiable launch process is an RTM, which serves as the trust anchor for the measurement process. Currently, TEEs use one of three types of RTMs, namely, static (SRTM), dynamic (DRTM), and hardware based (HW), as summarized in \cref{tab:attestation}. 

SRTM is created by an unbroken chain of trust from the code module that first executes at system reset to the code that runs in the \enclave. This chain of trust usually only includes all the software components of the \enclave{'s} TCB, as shown in \cref{fig:rtm}. The chain typically does not include the operating system. Such a solution can bootstrap the run-time components of the system TCB before any untrusted components (\advTEE, \advApp, \advSSW, and \advPer) are even active. For solutions that consider \advBus, such bootstrapping must ensure that no off-chip components (e.g., platform TPM) are required during this process. A typical SRTM could be implemented entirely in hardware or immutable software in a BootROM.

In contrast, solutions that use a DRTM can establish a new RTM without trusting the code executed prior to it since system reset (see \cref{fig:rtm:drtm}). So, these solutions must implement architectural extensions to protect the bootstrap process against adversaries (\advTEE, \advApp, \advSSW, and \advPer) that may potentially be active at the time of \enclave{} launch~\cite{mccune2008flicker, mccune2008low,deng2014equalvisor}. This can be done through specialized hardware instructions that first suspend all other active processes (hence, \advTEE, \advApp, \advSSW) and disable all IO devices and interrupts (hence, \advPer{}). Then, the hardware loads, measures, and authenticates a signed-code module that serves as the TCB of the actual \enclave{}. Once the hardware has verified the signed-code module and potentially recorded its measurement, it executes the module that in turn loads and measures the \enclave{} itself.

Most surveyed TEEs use SRTM (\cref{tab:attestation}). Only a few systems from two commercial processor manufacturers (Intel and AMD) leverage DRTM. We speculate that the main motivation of DRTM --- excluding boot code from the TCB --- only applies to platforms with a large amount of legacy boot code (e.g., x86 BIOS\cite{kauer2007oslo}).

\subsection{Measurement}
\label{sec:attest:measurement}
In SRTM and DRTM solutions, each entity in the chain of trust up to the \enclave, starting at the RTM, measures the next component before transferring execution control to it. In practice, all components in such a chain of trust are not only measured but also integrity-checked (e.g., by verifying a signature, checking the measurement against a reference value) before they are executed. We note that all surveyed TEEs use very similar techniques for measurement and we did not discover major differences. We refer the reader to \cref{sec:appendix:measurement} for a further discussion on implementation details of a typical measurement mechanism.

\subsection{Architectural Support for  Attestation}
Attestation is the third and final step of verifiable launch, where a verifier checks that the \enclave{} has been launched correctly and that its initial state is as expected. More specifically, the verifier ensures that the \enclave{}'s measurement and its underlying TCB match their expected reference values. There are two flavors of attestation: local attestation and remote attestation.
Local attestation is applicable when a verifier is co-located with the \enclave{} on the same platform. In contrast, remote attestation is meant for use by a remote verifier that is not on the same platform as the \enclave{} being attested. Remote attestation schemes usually rely on asymmetric cryptography and often incur the cost of checking one or more certificate chains. This can be expensive, especially when implemented in hardware. In contrast, local attestation is typically implemented using symmetric cryptography and tends to be more efficient. 
Most existing TEE solutions support remote attestation (\cref{tab:attestation}), but only a handful specify both local and remote attestation, as summarized in \cref{tab:attestation}. Note that the remote attestation of some TEEs can trivially be reused for local attestation.

\begin{table}
    \centering
    \begin{tabular}{@{} ll lllcc @{}} \toprule
            & \multirow{2}{*}[-3pt]{Name} & \multirow{2}{*}[-3pt]{ISA} & \multirow{2}{*}[-3pt]{RTM} & \multicolumn{2}{c}{Attestation} \\ \cmidrule{5-6}
             & & & & Local & Remote \\ \midrule
        \multirow{6}{*}{\begin{sideways}Industry\end{sideways}} 
            & Intel SGX~\cite{mckeen2013innovative,anati2013innovative}   & x86 & DRTM & \yes & \yes\\
            & Intel TDX~\cite{tdxbasespec}       & x86 & DRTM & \yes & \yes \\ 
            & AMD SEV-SNP\cite{kaplan2020amd}    & x86 & SRTM & \no & \yes\\
            & ARM TZ\cite{armtz}                 & ARM & SRTM & \no & \no \\
            & ARM Realms~\cite{armrealms}        & ARM & SRTM & \no & \yes \\ 
            & IBM PEF~\cite{hunt2021power}       & POWER & SRTM & \yes & \yes \\  
            \midrule
        \multirow{25}{*}{\begin{sideways}Academia\end{sideways}} 
            & Flicker~\cite{mccune2008flicker}          & x86 & DRTM & \no  & \yes \\
            & SEA~\cite{mccune2008low}                  & x86 & DRTM & \no  & \yes \\
            & SICE~\cite{azab2011sice}                  & x86 & SRTM & \no  & \yes \\
            & PodArch~\cite{shinde2015podarch}          & x86 & SRTM & \no  & \no \\
            & HyperCoffer~\cite{xia2013architecture}    & x86 & SRTM & \no  & \no \\
            & H-SVM~\cite{jin2011architectural,jin2015h}& x86 & SRTM & \no  & \no \\
            
            & EqualVisor~\cite{deng2014equalvisor}      & x86 & SRTM & \no  & \no \\
            & xu-cc15~\cite{xu2015architectural}        & x86 & SRTM & \no  & \no \\
            & wen-cf13~\cite{wen2013multi}              & x86 & SRTM & \no  & \no \\ 
            & Komodo~\cite{ferraiuolo2017komodo}        & ARM & SRTM & \no  & \yes \\
            & SANCTUARY~\cite{brasser2019sanctuary}     & ARM & SRTM & \yes & \yes \\
            & TrustICE~\cite{sun2015trustice}           & ARM & SRTM & \no  & \no \\
            & HA-VMSI~\cite{zhu2017ha}                  & ARM & SRTM & \no  & \yes \\
            & Sanctum~\cite{costan2016sanctum}          & RISC-V & SRTM & \yes & \yes \\
            & TIMBER-V~\cite{weiser2019timber}          & RISC-V & SRTM & \yes & \yes \\
            & Keystone~\cite{lee2020keystone}           & RISC-V & SRTM & \no  & \yes \\
            & Penglai~\cite{feng2021scalable}           & RISC-V & SRTM & \no  & \yes \\
            & CURE~\cite{bahami2021cure}                & RISC-V & SRTM & \no  & \yes \\
            & Iso-X~\cite{evtyushkin2014iso}            & OpenRISC & SRTM & \no  & \yes \\
            & HyperWall~\cite{szefer2012architectural}  & SPARC & SRTM & \no  & \yes \\
            & Sancus~\cite{noorman2013sancus,noorman2017sancus}  & MSP430 & HW & \no & \yes \\
            & TrustLite~\cite{koeberl2014trustlite}     & Custom & SRTM & \yes & \yes \\
            & TyTan~\cite{brasser2015tytan}             & Custom & SRTM & \yes & \yes \\
            & XOM~\cite{lie2000xom}                     & Custom & SRTM & \no  & \no \\
            & AEGIS~\cite{suh2003aegis}                 & Custom & SRTM & \no  & \no \\
        \bottomrule
    \end{tabular}
    \caption{The surveyed TEEs with their respective Root-of-Trust for Measurement (RTM) and their support for local and remote attestation. We use SRTM for static Root-of-Trust, DRTM for dynamic Root-of-Trust, and HW for hardware based systems that do not rely on SRTM or DRTM (c.f., \Cref{sec:attest:rtm}). \protect\yes{} indicates a TEE that describes a specific attestation mechanism whereas \protect\no{} is used for TEEs with no mention of such a mechanism. We note that some remote attestation schemes can be trivially re-used for local attestation.}
    \label{tab:attestation}
\end{table}

\subsection{Provisioning Secrets into an Enclave}
Provisioning secrets into \enclave{}s is often the last optional step during its launch. Some TEEs such as IBM PEF\cite{hunt2021power}, AMD SEV-SNP\cite{kaplan2020amd}, PodArch\cite{shinde2015podarch}, and Wen-cf13\cite{wen2013multi} allow \enclave{}s to be provisioned with secret data prior to the attestation. In this case, the \enclave{}'s initial state will contain some secret values also reflected in the measurement. This is achieved by an \enclave provisioning mechanism where the developer encrypts the secret data before delivering the \enclave{} binary to the platform. 

Other TEE designs require attestation before any secret data can be provisioned. To establish a communication channel bound to an attestation report, \enclave{}s in these TEEs may append some custom data (e.g., a public key certificate) to the attestation report\cite{costan2016intel,kaplan2020amd,lebedev2018sanctumattest}. Since this extra data is also authenticated during the attestation, its integrity is protected and can be leveraged to construct entire secure channels based on a key exchange protocol of choice. We note that AMD SEV\cite{kaplan2020amd} supports both the initial secret provisioning and establishing a secure channel bound to an attestation.

\section{Run-time Isolation}

\label{sec:isolation}

Following the setup of the \enclave, it begins executing. In order to prevent attackers from interfering with \enclave{} execution, all the resources belonging to the \enclave{}, including its CPU and memory, must be protected against unauthorized access and tampering. Such protection mechanisms are typically referred to as \emph{run-time isolation}. Below, we first describe a broad taxonomy of isolation strategies. Then, we discuss if and how one could apply them for CPU and memory isolation. Finally, we survey the set of strategies used by existing TEE designs to protect CPU and memory.

\subsection{Taxonomy of Isolation Strategies}
\label{sec:isol:tax}
\begin{figure}[tbp]
    \centering
    \begin{subfigure}[b]{0.33\linewidth}
        \centering
        \includegraphics[width=0.95\textwidth]{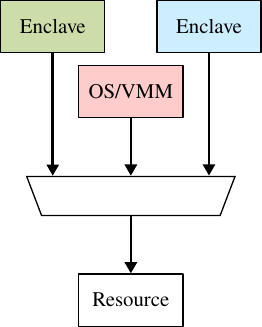}
        \caption{Temporal}
        \label{fig:isolstrat:temporal}
    \end{subfigure}\hfill
    \begin{subfigure}[b]{0.33\linewidth}
        \centering
        \includegraphics[width=0.95\textwidth]{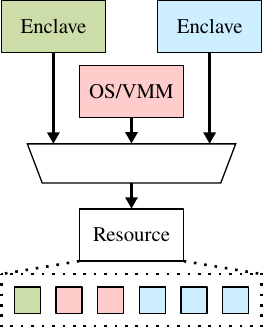}
        \caption{Spatio-temporal}
        \label{fig:isolstrat:spatiotemporal}
    \end{subfigure}\hfill
    \begin{subfigure}[b]{0.33\linewidth}
        \centering
        \includegraphics[width=0.95\textwidth]{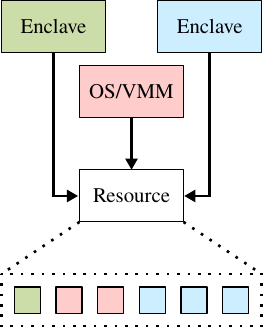}
        \caption{Spatial}
        \label{fig:isolstrat:spatial}
    \end{subfigure}
    \caption{Resources can be partitioned temporally, spatially, or a mix thereof (spatio-temporal).}
    \label{fig:isol_resource}
\end{figure}

\begin{figure}[tbp]
    \centering
    \begin{subfigure}[b]{0.38\linewidth}
        \centering
        \includegraphics[width=0.95\textwidth]{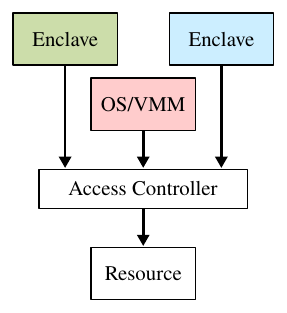}
        \caption{Logical}
        \label{fig:isolstrat:logical}
    \end{subfigure}
    \begin{subfigure}[b]{0.38\linewidth}
        \centering
        \includegraphics[width=0.95\textwidth]{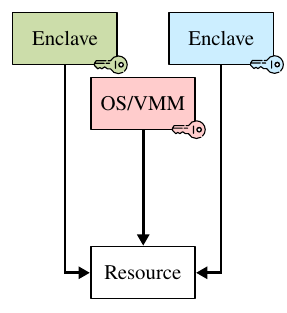}
        \caption{Cryptographic}
        \label{fig:isolstrat:crypto}
    \end{subfigure}
    \caption{Isolation enforcement strategies.}
    \label{fig:isol_types}
\end{figure}

In general, isolation mechanisms aim to achieve confidentiality and integrity of the protected resource, and can broadly be classified according to how resources are partitioned and how isolation is enforced. We describe these two dimensions below and depict them in \cref{fig:isol_types,fig:isol_resource}. 

\subsubsection*{Partitioning Resources}
Resources can be isolated in space, time, or a mix thereof (\cref{fig:isol_resource}). Note that this is not a categorical classification but rather a smooth range from fully temporal to fully spatial. As we will discuss later, these extreme cases of fully temporal or spatial only rarely appear. Instead, most isolation mechanism have some temporal and spatial aspects, i.e., they are spatio-temporal.

\emph{Temporal} partitioning splits a resource in the time domain, i.e., it securely multiplexes the same resource among multiple execution contexts over time. At any point in time, a single execution context has exclusive access to the resource. Temporal partitioning requires mechanisms to securely switch contexts while re-assigning the resource to a new execution context. Such secure context switches should be fast to not impact the general system performance. Temporal isolation is often used for resources that are costly to spatially partition and situations where concurrent access from multiple execution contexts to the same resource is not required.

In \emph{spatial} partitioning, resources are split such that trusted and untrusted contexts use separate, dedicated partitions. 
It can be used when there are multiple identical copies of the same resource (e.g., logical processor) or when the resource can be split into smaller identical copies. Note that a given execution context may be assigned more than one instance of a resource (e.g., multiple logical processors) if required. As a result, spatial isolation  techniques are often used for resources that are relatively cheap to replicate or to split. It also entails implementing mechanisms to ensure that the different copies of the resource can be used concurrently by different entities without any interference among them. 

\emph{Spatio-temporal} partitioning leverages both temporal as well as spatial aspects to partition a resource, e.g., the resource can be spatially partitioned but these partitions may change over time. This concept can only be used in resources that support both temporal and spatial partitioning. However, it may provide more flexibility and some performance advantages, and thus, is quite a popular choice. 

\subsubsection*{Enforcement}
In contrast to resource partitioning, enforcement of isolation strategies can be classified into two distinct categories: logical and cryptographic isolation. An overview of the two strategies is depicted in \cref{fig:isol_types}.

\emph{Logical isolation} leverages logical access control mechanisms to enforce isolation. These mechanism prohibit the adversary from accessing protected data. For example, a trusted context switch routine uses logical isolation to make sure the next execution thread cannot access any protected data by saving and purging the processor registers. On the other hand, many logical isolation mechanisms intercept data accesses and check the requests against some access control information. This access control information must be generated and managed by a trusted entity in the system, and it could be modified at run-time to enable flexible resource re-allocation. The resources (e.g., storage) needed to maintain the access control information varies depending on the granularity of the access control information and the type of resource being managed. Furthermore, the access control information itself must be protected against attacks.

As the name indicates, \emph{cryptographic isolation} uses cryptography to achieve isolation. Confidentiality is usually achieved via encryption; only authorized contexts with access to the correct cryptographic key material can decrypt data correctly. In contrast to logical isolation where protected data is not accessible at all, unauthorized contexts may read the ciphertext but they cannot retrieve the plaintext. Integrity is in part achieved through a cryptographic Message Authentication Code (MAC) stored alongside the data. This prevents an unauthorized context from tampering with data that does not belong to it because it cannot generate the correct MAC for a given piece of data without access to the correct keys. Achieving complete integrity with cryptographic isolation requires using anti-replay schemes that prevent re-injection of old data (with the correct, corresponding MAC) at a later point in time.

Below, we discuss the application of the above isolation strategies to CPU and memory and the resulting trade-offs.

\subsection{CPU Isolation}
While a typical CPU has many components, the discussion below focuses on the architectural state within the CPU such as the register state. This is mainly because many details on micro-architectural CPU state (e.g., intermediate CPU buffers, schedulers) are not publicly available for commercial TEE solutions. Even academic TEE solutions often omit these details. However, the impact of TEE implementations on memory-related micro-architectural structures (e.g., caches, TLBs) in the CPU are available and are analyzed in \Cref{sec:iso:memory}.

\subsubsection*{Choice of CPU Isolation Strategy}
While all isolation strategies are applicable to the architectural state within the CPU, most of them have considerable downsides, e.g., spatially reserving a (virtual) core exclusively for an \enclave{} incurs sub-optimal resource utilization and limits the number of concurrent \enclave{}s. Similarly, spatio-temporal approaches require extra hardware in a performance-critical part of the processor to protect spatially separated data. Hence, these techniques are not well suited for CPU isolation. In contrast, temporal partitioning is very well suited for CPU isolation as it does not add additional runtime checks beside a trusted context switch.

On the enforcement side, cryptographic enforcement suffers from a large performance overhead due to the extra cryptographic hardware on the fast-path of the processor. On the other hand, temporal partitioning combined with logical enforcement does not exhibit such overheads besides requiring a fast and secure context switching routine that temporally separates multiple execution contexts on the same CPU thread. 

In-line with our analysis, all existing TEE solutions that we studied use temporal partitioning combined with logical enforcement for CPU state. Our results are listed in \cref{tab:cpuisol}.

\begin{figure*}
    \centering
    \includegraphics[width=0.7\linewidth]{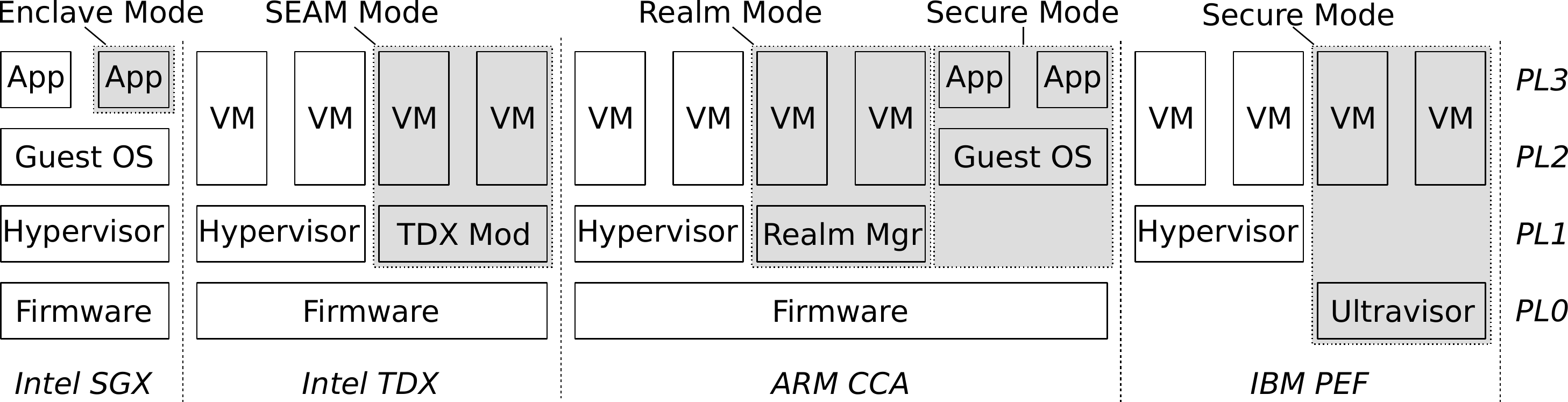}
    \caption{Some TEEs introduce new CPU modes with their design, e.g., Intel SGX, Intel TDX, ARM CCA, and IBM PEF.}
    \label{fig:new_privlevels}
\end{figure*}

\subsubsection*{Architectural Support for CPU Isolation}
As mentioned before, temporal partitioning with logical enforcement is typically implemented through a secure context switch routine that saves, purges, and restores the execution contexts. 
The context switch routine must ensure that the data from an \enclave{} does not leak to any untrusted context or another \enclave{} that follows it in the execution schedule. In addition, an untrusted context must not be able to tamper with or, more generally, control the CPU execution state of an \enclave{} when it is being started or resumed. While there are multiple options to save and restore execution contexts such as encrypting the registers on \enclave{} exit, most TEEs save the context to the \enclave{}'s private memory and afterwards purge the register values. To achieve this, TEEs rely on their TCB for setting up the CPU state correctly before starting an \enclave{} and scrubbing the CPU state while exiting an \enclave. %

In order to ensure that the TCB can fully mediate every context switch, TEEs leverage multiple CPU modes, privilege levels, and in some cases, a combination thereof. This ensures that all transitions into and out of an \enclave{} occur from TCB-controlled mode(s)/privilege level(s). The actual solution used by a TEE design often depends on the underlying processor's instruction set architecture. 
Commercial processors from Intel, AMD, ARM, and IBM often add new execution modes to support TEEs (see \cref{fig:new_privlevels}). %
In contrast, most academic TEEs do not introduce any new processor modes or privilege levels. Instead, they rely on the firmware running in an existing high-privilege level (PL0) for secure context switching. There are a few proposals in which the hardware itself facilitates the temporal isolation during the context switch~\cite{evtyushkin2014iso,jin2015h,szefer2012architectural,xu2015architectural,wen2013multi,suh2003aegis,lie2000xom}. %

Besides context switching, CPU modes and privilege levels are often necessary for securely running the \enclave{} and its software TCB components (if any), as summarized in \cref{tab:cpuisol}. While the \enclave{} code typically encompasses an application or a virtual machine and runs at lower privilege (PL2, PL3), the TCB tends to run at higher privilege in most TEE designs (PL0, PL1). This allows the TCB to  implement many other types of security mechanisms such as measurement and attestation as discussed in \Cref{sec:attestation} as well as isolation of memory, trusted IO, and secure storage as discussed in the rest of this paper.

\begin{table}
    \centering
    \captionsetup{width=.9\linewidth}
    \begin{threeparttable}
    
    \begin{tabular}{ll ccc}\toprule
            & \multirow{2}{*}[-3pt]{Name} & \multirow{2}{*}[-3pt]{\makecell{Isol\\Strat}} & \multicolumn{2}{c}{Privilege Level} \\ \cmidrule{4-5}
            & & & Enclave & Software TCB \\ \midrule
        \multirow{6}{*}{\begin{sideways}Industry\end{sideways}} 
            & Intel SGX~\cite{mckeen2013innovative,anati2013innovative}  & T-L & App & - \\
            & Intel TDX~\cite{tdxbasespec}      & T-L & VM  & PL1 \\ 
            & AMD SEV-SNP\cite{kaplan2020amd}   & T-L & VM & -$^\dagger{}$ \\
            & ARM TZ\cite{armtz}                & T-L & App/VM & PL0+(PL1/2)$^\ddagger{}$ \\
            & ARM CCA~\cite{armrealms}          & T-L & VM & PL0+(PL1)$^\ddagger{}$ \\ 
            & IBM PEF~\cite{hunt2021power}      & T-L & VM & PL0 \\  
            \midrule
        \multirow{25}{*}{\begin{sideways}Academia\end{sideways}} 
            & Flicker~\cite{mccune2008flicker}      & T-L & VM & - \\
            & SEA~\cite{mccune2008low}              & T-L & VM & - \\
            & SICE~\cite{azab2011sice}              & T-L & VM & PL0 \\
            & PodArch~\cite{shinde2015podarch}      & T-L & App & - \\
            & HyperCoffer~\cite{xia2013architecture} & T-L & VM & PL0 \\
            & H-SVM~\cite{jin2011architectural,jin2015h} & T-L & VM & - \\
            & EqualVisor~\cite{deng2014equalvisor}  & T-L & VM & PL1 \\
            & xu-cc15~\cite{xu2015architectural}    & ? & App & - \\
            & wen-cf13~\cite{wen2013multi}          & T-L & VM & - \\
            
            & Komodo~\cite{ferraiuolo2017komodo}    & T-L & App & PL0 + PL2 \\
            & SANCTUARY~\cite{brasser2019sanctuary} & T-L & App & PL0 + PL2 \\
            & TrustICE~\cite{sun2015trustice}       & T-L & App & PL0 + PL2 \\
            & HA-VMSI~\cite{zhu2017ha}              & T-L & VM & PL0 \\
            & Sanctum~\cite{costan2016sanctum}      & T-L & App & PL0 \\
            & TIMBER-V~\cite{weiser2019timber}      & T-L & App & PL0 \\
            & Keystone~\cite{lee2020keystone}       & T-L & App & PL0 \\
            & Penglai~\cite{feng2021scalable}       & T-L & App & PL0 \\
            & CURE~\cite{bahami2021cure}            & T-L & App/VM & PL0 \\
            & Iso-X~\cite{evtyushkin2014iso}        & T-L & App & - \\
            & HyperWall~\cite{szefer2012architectural} & T-L & VM & -  \\
            & Sancus~\cite{noorman2013sancus,noorman2017sancus} & T-L & App & - \\
            & TrustLite~\cite{koeberl2014trustlite} & T-L & App & PL0 \\
            & TyTan~\cite{brasser2015tytan}         & T-L & App & PL0 \\
            & XOM~\cite{lie2000xom}                 & T-L & App & - \\
            & AEGIS~\cite{suh2003aegis}             & T-L & App & - \\
        \bottomrule
    \end{tabular}
    \begin{tablenotes}
        \item [$\dagger$] AMD SEV-SNP palces the TCB in separate co-processor\cite{kaplan2020amd}.
        \item [$\ddagger{}$] ARM TZ and ARM Realms only provide the hardware primitives to implement a TEE. There are multiple options to implement the software TCB in different privilege levels.
    \end{tablenotes}
    \end{threeparttable}
    \caption{Summary of CPU Isolation in TEEs: All TEE solutions use temporal and logical isolation (indicated by T-L) to securely share the CPU among execution contexts. \Enclave{}s are run in either an App (PL3) or as a VM (PL2). The software TCB is implemented in a more privileged level than the \enclave.}
    \label{tab:cpuisol}
\end{table}

\subsection{Memory Isolation}
\label{sec:iso:memory}
Ensuring that the memory used by an \enclave{} is protected at run-time against unauthorized access and modification is a particularly challenging aspect of TEE design. Such protections must cover not only the actual off-chip memory, but also any code/data that resides in the on-chip micro-architectural structures such as instruction and data caches. Furthermore, since most TEEs support virtual memory, trustworthiness (or lack thereof) of the translation structures such as page tables that convert virtual addresses to physical addresses are a key design aspect of all memory isolation solutions. Similar to processor caches, translation look-aside buffers (TLB) holding recent page translations must also be protected against misuse and misconfiguration.

\subsubsection*{Choice of Memory Isolation Strategy}
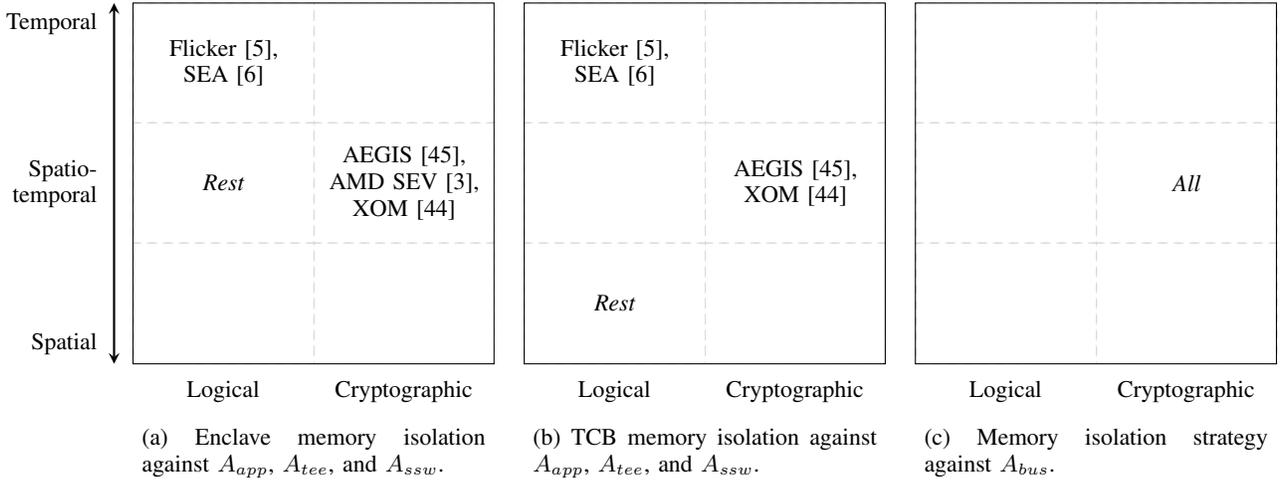
\begin{figure*}[tbp]
    \centering
    \def\h{1.6cm} %
\def\w{2.4cm} %
\def\d{0.5cm} %

\newcommand*{\drawenv}{%
    \coordinate (temporal) at (0, 2.5*\h);
    \coordinate (spatial) at (0, 0.5*\h);
    \coordinate (spatemp) at (0, 1.5*\h);
    \coordinate (logical) at (\w/2, 0);
    \coordinate (crypto) at (1.5*\w, 0);

    \node[draw=none, align=center, anchor=north] at (logical |- 0,-0.1) {Logical};
    \node[draw=none, align=center, anchor=north] at (crypto |- 0,-0.1) {Cryptographic};
}
\begin{subfigure}[T]{0.085\textwidth}
    \centering
        \begin{tikzpicture}[align=center,>=stealth]
            \coordinate (temporal) at (0, 2.5*\h);
            \coordinate (spatial) at (0, 0.5*\h);
            \coordinate (spatemp) at (0, 1.5*\h);
            \coordinate (logical) at (\w/2, 0);
            \coordinate (crypto) at (1.5*\w, 0);
            \path[draw=black,thick,<->] (0,0) -- (0, 3*\h);
            \node[draw=none, align=right, anchor=south east] at ($(spatial) + (-0.1,-0.5*\h)$) {Spatial};
            \node[draw=none, align=right, anchor=north east] at ($(temporal) + (-0.1,0.5*\h)$) {Temporal};
            \node[draw=none, align=right, anchor=east] at (spatemp  -| -0.1,0) {Spatio-\\temporal};
            
        \end{tikzpicture}
\end{subfigure}
\begin{subfigure}[T]{0.28\textwidth}
    \centering\captionsetup{width=0.9\linewidth}
        \begin{tikzpicture}[align=center,-stealth]
            \tikzset{>=latex}
            \tikzset{cat/.style={draw=black!15,dashed,minimum width=\w, minimum height=\h,text width=0.9*\w,anchor=south west}}

            \drawenv

            \node[cat] at (0, 0) {};
            \node[cat] at (0, \h) {\textit{Rest}};
            \node[cat] at (0, 2*\h) {Flicker~\cite{mccune2008flicker}, SEA~\cite{mccune2008low}};
            \node[cat] at (\w, 0) {};
            \node[cat] at (\w, \h) {AEGIS~\cite{suh2003aegis}, AMD SEV~\cite{kaplan2020amd}, XOM~\cite{lie2000xom}};
            \node[cat] at (\w, 2*\h) {};

            \path[draw=black, -] (0, 0) -- (0, 3*\h) -- (2*\w, 3*\h) -- (2*\w, 0) -- cycle;
        \end{tikzpicture}
    \caption{Enclave memory isolation against \advApp, \advTEE, and \advSSW.}
    \label{fig:isol:mem:cpu}
\end{subfigure}
\begin{subfigure}[T]{0.28\textwidth}
    \centering\captionsetup{width=0.9\linewidth}
    \begin{tikzpicture}[align=center,-stealth]
        \tikzset{>=latex}
        \tikzset{cat/.style={draw=black!15,dashed,minimum width=\w, minimum height=\h,text width=0.9*\w,anchor=south west}}

        \drawenv

        \node[cat] at (0, 0) {\textit{Rest}};
        \node[cat] at (0, \h) {};
        \node[cat] at (0, 2*\h) {Flicker~\cite{mccune2008flicker}, SEA~\cite{mccune2008low}};
        \node[cat] at (\w, 0) {};
        \node[cat] at (\w, \h) {AEGIS~\cite{suh2003aegis}, XOM~\cite{lie2000xom}};
        \node[cat] at (\w, 2*\h) {};

        \path[draw=black, -] (0, 0) -- (0, 3*\h) -- (2*\w, 3*\h) -- (2*\w, 0) -- cycle;
    \end{tikzpicture}
    \caption{TCB memory isolation against \advApp, \advTEE, and \advSSW.}
    \label{fig:isol:mem:cache}
\end{subfigure}
\begin{subfigure}[T]{0.28\textwidth}
    \centering\captionsetup{width=0.9\linewidth}
    \begin{tikzpicture}[align=center,-stealth]
        \tikzset{>=latex}
        \tikzset{cat/.style={draw=black!15,dashed,minimum width=\w, minimum height=\h,text width=0.9*\w,anchor=south west}}

        \drawenv

        \node[cat] at (0, 0) {};
        \node[cat] at (0, \h) {};
        \node[cat] at (0, 2*\h) {};
        \node[cat] at (\w, 0) {};
        \node[cat] at (\w, \h) {\textit{All}};
        \node[cat] at (\w, 2*\h) {};

        \path[draw=black, -] (0, 0) -- (0, 3*\h) -- (2*\w, 3*\h) -- (2*\w, 0) -- cycle;
    \end{tikzpicture}
    \caption{Memory isolation strategy against \advBus.}
    \label{fig:isol:mem:mem}
\end{subfigure}
    \caption{The isolation strategies employed in main memory according to the adversaries they protect against. Note that many TEEs use distinct strategies for different adversaries. Not all surveyed TEEs support a physical adversary (c.f., \cref{tab:appendix:memoryisol} in the appendix).}
    \label{fig:isol:memory}
\end{figure*}

All previously discussed isolation strategies can be applied to memory and we discuss the implications of these strategies below. While all TEE designs used temporal partitioning for CPU isolation, their memory isolation strategies are diverse. In fact, many TEE designs use different strategies based on the type(s) of attacker(s) under consideration, as shown in \cref{fig:isol:memory}.

Full spatial partitioning of memory implies reserving one or more memory regions for exclusive use by an \enclave{} or the TCB and these regions remain assigned to it until the next system reboot. Spatial partitioning works well when the number of \enclave{s} that the system must support is small, and their memory resource requirements are fairly static and well-known in advance. It also works well for coarse-grained memory protections that protect access control information used for logical isolation (as explained below).

Fully temporal memory partitioning involves allowing memory accesses only from the currently active execution context and securely saving/restoring memory content during context switches. Its use for memory isolation is limited to scenarios where only a single execution context is active at any point in time. Hence, it is not efficient for TEEs supporting concurrent \enclave{s}. Besides, saving memory content to disk can take considerable time. Thus, temporal memory partitioning is rarely used.

Spatio-temporal memory partitioning is the preferred style due to its flexibility: memory regions can be re-allocated to a different execution context over time. Hence, it is useful in systems where memory requirements for \enclave{s} cannot be predicted upfront.

Logical enforcement relies on access control mechanism to only allow authorized accesses to \enclave{} resources. Logical isolation requires every access to be checked against access control information, e.g., by a memory management unit (MMU). 
Furthermore, achieving integrity protection against software adversaries is rather efficient, with only a small amount of access control information per \enclave{}. 

Cryptographic enforcement achieves confidentiality through encryption. It ensures integrity by maintaining a cryptographic MAC for each block of memory of configurable size. Protection against replay attacks is achieved by maintaining freshness information (e.g., counters), often in the form of a Merkle tree\cite{suh2003efficient,suh2003aegis}. In contrast to all the isolation strategies above, cryptographic memory isolation can protect against \advBus{}. However, cryptographic isolation is hard to scale, especially if different execution contexts need separate keys, because it requires maintaining large amounts of cryptographic keying material on-die within the SoC. Furthermore, achieving integrity and anti-replay properties using cryptographic isolation results in storage overheads (e.g., for the MAC and anti-replay metadata) as well as latency and throughput overheads\cite{suh2003efficient}.

In the surveyed TEEs, we found a variety of such strategies. While there often is a preferred design choice (indicated by \emph{Rest} or \emph{all} in \cref{fig:isol:memory}), other options exist and seem to be practical. We highlight that many TEEs use multiple strategies simultaneously, e.g., Intel SGX\cite{mckeen2013innovative} uses spatio-temporal partitioning and logical enforcement to protect enclave memory, but it uses purely spatial partitioning to protect its TCB, and cryptographic isolation to cope with \advBus{}.

\subsubsection*{Architectural Support for Memory Isolation}
Logical enforcement of spatial or spatio-temporal memory isolation is facilitated by an access control check. All surveyed TEEs use one of two options for the access control check: memory protection units (MPU) or memory management units (MMU). We refer the reader to \cref{sec:appendix:mpu} for a discussion on the implementation details of these two options. One of the main differences between MPU and MMU based enforcement, is that the former operates on physical addresses and the latter on virtual addresses. Also, MPUs typically only support a limited number of rules, whereas MMUs are more flexible. We note that TEEs that leverage MMUs are typically more complex and often come with an in-depth security analysis. On the other hand, MPUs are rather simple and thus may simplify the security analysis. Many modern academic TEEs rely on an MPU to provide isolation\cite{lee2020keystone,bahami2021cure,weiser2019timber,brasser2019sanctuary,costan2016sanctum}. On the other hand, many commercial TEEs seem to appreciate the increased flexibility of the MMU\cite{mckeen2013innovative,kaplan2020amd,hunt2021power,armrealms}.

We note that the access control information used to enforce memory isolation, i.e., trusted page tables for an MMU, access control rules for the MPU, or other secondary metadata, must itself be protected against unauthorized access. This is done typically through spatial partitioning, i.e., memory for such structures is allocated at boot and protected through a simplified MPU (e.g., range registers).

\subsubsection*{Caches}
\begin{figure*}[tbp]
    \centering
    \def\h{1.6cm} %
\def\w{2.4cm} %
\def\d{0.5cm} %

\newcommand*{\drawenv}{%
    \coordinate (temporal) at (0, 2.5*\h);
    \coordinate (spatial) at (0, 0.5*\h);
    \coordinate (spatemp) at (0, 1.5*\h);
    \coordinate (logical) at (\w/2, 0);
    \coordinate (crypto) at (1.5*\w, 0);

    \node[draw=none, align=center, anchor=north] at (logical |- 0,-0.1) {Logical};
    \node[draw=none, align=center, anchor=north] at (crypto |- 0,-0.1) {Cryptographic};
}
\begin{subfigure}[T]{0.095\textwidth}
    \centering
        \begin{tikzpicture}[align=center,>=stealth]
            \coordinate (temporal) at (0, 2.5*\h);
            \coordinate (spatial) at (0, 0.5*\h);
            \coordinate (spatemp) at (0, 1.5*\h);
            \coordinate (logical) at (\w/2, 0);
            \coordinate (crypto) at (1.5*\w, 0);
            \path[draw=black,thick,<->] (0,0) -- (0, 3*\h);
            \node[draw=none, align=right, anchor=south east] at ($(spatial) + (-0.1,-0.5*\h)$) {Spatial};
            \node[draw=none, align=right, anchor=north east] at ($(temporal) + (-0.1,0.5*\h)$) {Temporal};
            \node[draw=none, align=right, anchor=east] at (spatemp  -| -0.1,0) {Spatio-\\temporal};
            
        \end{tikzpicture}
\end{subfigure}
\begin{subfigure}[T]{0.28\textwidth}
    \centering
        \begin{tikzpicture}[align=center,-stealth]
            \tikzset{>=latex}
            \tikzset{cat/.style={draw=black!15,dashed,minimum width=\w, minimum height=\h,text width=0.9*\w,anchor=south west}}

            \drawenv

            \node[cat] at (0, 0) {Penglai~\cite{feng2021scalable}};
            \node[cat] at (0, \h) {\textit{Rest}};
            \node[cat] at (0, 2*\h) {CURE~\cite{bahami2021cure}, Flicker~\cite{mccune2008flicker}, Sanctum~\cite{costan2016sanctum}, SEA~\cite{mccune2008low}};
            \node[cat] at (\w, 0) {};
            \node[cat] at (\w, \h) {};
            \node[cat] at (\w, 2*\h) {};

            \path[draw=black, -] (0, 0) -- (0, 3*\h) -- (2*\w, 3*\h) -- (2*\w, 0) -- cycle;
        \end{tikzpicture}
    \caption{Local Cache}
    \label{fig:isol:cpu}
\end{subfigure}
\begin{subfigure}[T]{0.28\textwidth}
    \centering
    \begin{tikzpicture}[align=center,-stealth]
        \tikzset{>=latex}
        \tikzset{cat/.style={draw=black!15,dashed,minimum width=\w, minimum height=\h,text width=0.9*\w,anchor=south west}}

        \drawenv

        \node[cat] at (0, 0) {CURE~\cite{bahami2021cure}, Keystone~\cite{lee2020keystone}, Penglai~\cite{feng2021scalable}, Sanctum~\cite{costan2016sanctum}};
        \node[cat] at (0, \h) {\textit{Rest}};
        \node[cat] at (0, 2*\h) {Flicker~\cite{mccune2008flicker},\\SEA~\cite{mccune2008low}};
        \node[cat] at (\w, 0) {};
        \node[cat] at (\w, \h) {};
        \node[cat] at (\w, 2*\h) {};

        \path[draw=black, -] (0, 0) -- (0, 3*\h) -- (2*\w, 3*\h) -- (2*\w, 0) -- cycle;
    \end{tikzpicture}
    \caption{Shared Cache}
    \label{fig:isol:sharedcache}
\end{subfigure}
    \caption{The isolation strategies employed in local and shared caches against software adversaries (\advApp, \advTEE, and \advSSW). Most TEEs use spatio-temporal logical isolation. However, some TEEs leverage fully temporal or spatial partitioning.}
    \label{fig:isol:cache}
\end{figure*}
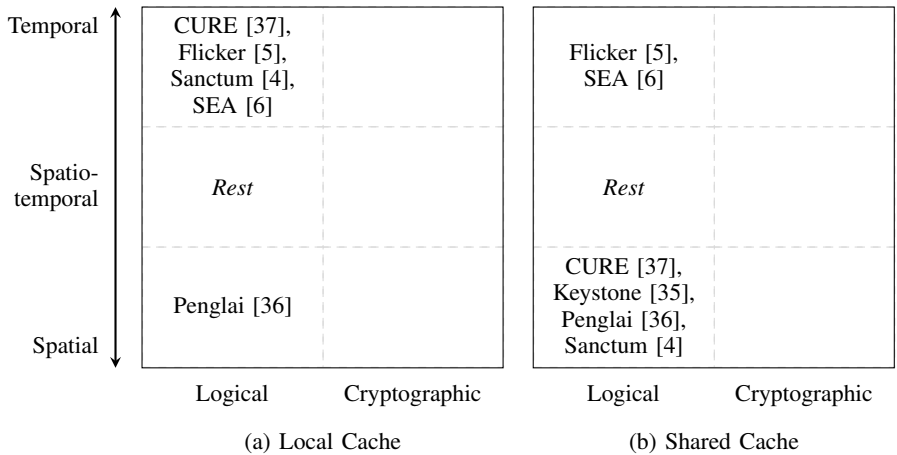

Caches contain recently used portions of memory for improved software performance. Often, CPUs contain multiple cache layers, some exclusive to a core and others that are shared. In some TEE architectures, the MPU/MMU in the CPU and their IO counterparts prevent untrusted entities from accessing \enclave{} data in the caches (e.g., Intel SGX). In other TEEs where such accesses cannot be prevented by existing mechanisms, additional cache protection mechanisms isolate \enclave{} data in the caches (e.g., Arm TrustZone). A summary of the isolation strategies for caches is depicted in \cref{fig:isol:cache}.

Spatial cache partitioning, i.e., reserving portions of the cache for exclusive use by an \enclave{}, does not scale well, reduces resource utilization, and can lead to potential performance degradation. However, it can be used to mitigate side-channel attacks\cite{costan2016sanctum,bourgeat2019mi6,lee2020keystone,feng2021scalable,bahami2021cure}. Temporal cache partitioning is not very efficient because it requires flushing the entire cache on every transition among execution contexts. Hence, it is only used by a handful of TEE designs to protect against side-channel leakage and is limited to small caches that are exclusive to a single core~\cite{costan2016sanctum,brasser2019sanctuary,lee2020keystone}. Cryptographic enforcement is not suitable for micro-architectural structures like caches because of the cost due to extra cryptographic hardware as well as its limited latency and throughput. So, most caches in TEE solutions today implement spatio-temporal and logical cache isolation.

\section{Trusted IO}
\label{sec:io}
While early TEE designs only focus on the CPU, recent interest in secure interactions with external devices has inspired multiple approaches for trusted IO\cite{li2018vbutton, ying2018truz,bahami2021cure,nasahl2021hector}. Trusted IO has two components: \emph{(i)} confidentiality and integrity for the \enclave{}'s accesses to the device, i.e., establishing a \emph{trusted path}, and \emph{(ii)} protecting \enclave data on the device through a \emph{trusted device architecture}. We discuss existing solutions for each of these components below.

\subsection{Establishing a Trusted Path} 
Originally, the term \emph{trusted path} was used in the context of enabling trusted IO interactions for users\cite{zhou2012building}. However, today, with the increasing diversity of IO devices and the trend towards enabling their use with TEEs, a trusted path is also used to refer to the (secure) communication channel between an \enclave{} and a device. 

Both logical and cryptographic isolation techniques can be used to establish a trusted path. If the trusted path has multiple hops, each hop could implement a different type of isolation. While both logical and cryptographic trusted paths can protect against \advTEE, \advApp, \advSSW, \advBoot and \advPer, only cryptographic trusted paths can protect against \advBus. 
Below, we discuss different ways to implement logical and cryptographic trusted paths.

\subsubsection*{Architectural Support for Logical Trusted Path}
Logical trusted paths require architectural support to enable two types of IO interactions: direct-memory-access (DMA) and memory mapped IO (MMIO). In this discussion, we focus on MMIO and refer the reader to \Cref{sec:iso:memory} for a detailed analysis of isolation mechanisms for DMA.

One way to build a logical trusted path for MMIO is through access control filters that allow/deny accesses based on the origin or the destination of the MMIO request. 
Such filters could be static or programmable at run-time by a trusted entity. Many systems rely on ARM's TrustZone Protection Controller to filter accesses to peripherals\cite{li2018vbutton, ying2018truz, li2015adattester, li2014building}. CURE~\cite{bahami2021cure}, TrustOTP~\cite{sun2015trustotp} and HectorV~\cite{nasahl2021hector} rely on similar but more complex filters in front of each peripheral to allow/disallow accesses. 
Another option to build a logical trusted path is through trusted memory mappings via secure MPU configurations (e.g., TrustLite~\cite{koeberl2014trustlite}) or related metadata structures (e.g., HIX~\cite{jang2019heterogeneous}) that are checked every time an \enclave{} tries to access a device.

\subsubsection*{Architectural Support for Cryptographic Trusted Path}
End-to-end cryptographic trusted paths rely on a secure channel established between two endpoints: the \enclave{} and the device. This approach incurs overhead related to the cryptographic operations and hardware. To establish a secure channel, the device and \enclave{} must be provisioned with credentials and cryptographic keys to use for authentication and optionally, attestation.
Examples of solutions that use a cryptographic trusted path include  Fidelius~\cite{eskandarian2019fidelius} and HIX~\cite{jang2019heterogeneous}. Sometimes, cryptographic channels may be multi-hop, i.e., include a trusted intermediate hardware component between the \enclave{} and the device (e.g., Bastion SGX~\cite{peters2018bastion}, ProtectIOn~\cite{dhar2019protection}, and HETEE~\cite{zhu2019enabling}). Cryptographic trusted paths at the link level (as opposed to high software levels) are emerging to protect specifically against \advBus{}\cite{synopsysPCIencryption}.

Certain trusted path solutions combine both cryptographic and logical trusted paths. For example, SGXIO~\cite{weiser2017sgxio} uses a cryptographic path between the CPU package and the device, but isolates accesses from different \enclave{s} using a trusted intermediary on the CPU package with exclusive access to the device. It is also possible to use a different type of trusted path for the DMA and MMIO respectively; HIX\cite{jang2019heterogeneous} uses a logical trusted path for MMIO and a cryptographic trusted path for DMA.

\subsection{Trusted Device Architectures} 
For certain types of trusted IO usages, it is sufficient to establish a trusted path from the \enclave{} to the device. Examples include peripherals that do not  process user data in clear text (e.g., encrypted storage devices, network cards). However, newly emerging  devices such as accelerators are used for computations on user data. Such accelerators must ensure the confidentiality and integrity of every \enclave{}'s data just as the CPU does. 

Today, a variety of accelerators exist: e.g., custom chips for AI processing, Field Programmable Logic Arrays (FPGAs) and general purpose Graphics Processing Units (GPUs). These systems differ greatly in terms of their underlying architectures and hence, the exact mechanisms that they must implement to isolate \enclave{} data also varies. The details of these mechanisms for individual accelerators are out-of-scope for this paper. However, there are still a set of high-level isolation techniques based on the strategies discussed in \Cref{sec:isol:tax} that apply to many of these accelerators as discussed  below. 

\subsubsection*{Spatial Partitioning} Assigning separate/dedicated instances of an accelerator to each \enclave{} for the entire lifetime of the platform is typically not resource or cost efficient. Therefore, while it is feasible to do in theory, spatial isolation is not very practical. However, if such dedicated instances are available, then establishing a trusted path to the device suffices and no additional device requirements arise.

\subsubsection*{Temporal Partitioning} Until recently, accelerators were built assuming exclusive use by a single execution context at any given time. Therefore, temporal partitioning, i.e., sharing it among multiple contexts over time, is a common approach. Such temporal partitioning requires a secure context switching mechanism which could be implemented either in software or by enhancing the accelerator hardware itself. Example solutions that rely on such temporal partitioning include HETEE~\cite{zhu2019enabling} for generic single-use accelerators, ZeroKernel~\cite{kwon2019zerokernel} and HIX\cite{jang2019heterogeneous} for GPUs, as well as MeetGo~\cite{oh2021meetgo} and ShEF~\cite{zhao2021shef} for FPGAs.
    
\subsubsection*{Spatio-temporal Partitioning and Logical Enforcement} This is typically used when multiple \enclave{s} need concurrent access to an accelerator due to its flexibility. Such spatio-temporal isolation requires hardware support by the device architecture to enable true multi-tenancy. Again, the exact set of enhancements are device-specific, but they often involve maintaining access control information to track ownership of resources, i.e., mapping of resources to \enclave{}s. Examples of such solutions include Graviton~\cite{volos2018graviton}, Telekine~\cite{hunt2020telekine} and SEGIVE~\cite{wang2020segive} for GPUs as well as Trustore~\cite{oh2020trustore} for FPGAs. 
    
\subsubsection*{Cryptographic Enforcement} This is not well-suited for protecting on-chip accelerator resources (e.g., caches, TLBs) for the same reasons as it is not optimal for protecting on-chip CPU resources, namely, due to the performance overhead as well as additional area cost for all the cryptography hardware. However, cryptographic enforcement can be applied specifically to device-side memory resources just like they apply to DRAM on the CPU as described in \Cref{sec:isolation}.

\section{Secure Storage}

In many applications, \enclave{}s are required to retain certain (persistent) state across different invocations. The process of protecting the data in this way through encryption is referred to as \emph{sealing} and the reverse (decryption) process that accounts for \enclave{} state is called \emph{unsealing}. 

While such secure storage is a very common requirement, it is not described explicitly by most existing TEE designs. Some TEEs support primitives that can be leveraged for secure storage (e.g., AMD SEV-SNP\cite{kaplan2020amd}), but do not describe a full solution; therefore, we do not discuss them any further here. So, in the rest of this section, we focus on the TEEs that provide a complete description of their support for sealing: Flicker~\cite{mccune2008flicker}, SEA~\cite{mccune2008low}, IBM-PEF~\cite{hunt2021power}, Intel SGX\cite{anati2013innovative}, TIMBERV~\cite{weiser2019timber}, Keystone\cite{lee2020keystone}, and Sanctuary~\cite{brasser2019sanctuary}.

\subsection{Sealing Solutions and Trade-offs}
All sealing proposals in the surveyed TEEs closely resemble the original proposal based on the TPM~\cite{tpm}. Flicker~\cite{mccune2008flicker}, SEA~\cite{mccune2008low} and IBM-PEF~\cite{hunt2021power} directly rely on the original sealing mechanism of a TPM. This typically involves generating an asymmetric key pair and using it to encrypt a secret such that it can be decrypted (unsealed) successfully only when the system configuration matches the one at the time of encryption (sealing). The system configuration information used during the sealing and unsealing process with a TPM uses the measurements recorded in the TPM's Platform Configuration Registers (PCRs). Since TPMs have limited storage, a typical way to protect large amounts of data is to generate a symmetric key for bulk data encryption and then, seal that key to a TPM. 

There exist many different forms of establishing which \enclave{} can unseal previously sealed data: Some TEEs only allow an \enclave{} with the same measurement and on the same platform with a specific TCB version to unseal the data\cite{weiser2019timber,brasser2019sanctuary, kaplan2020amd}. Other proposals allow all \enclave{}s signed by the same developer to unseal each other's data\cite{anati2013innovative,epidIntel2016}. Some cloud TEEs also allow \enclave{}s to come with a migration policy to migrate sealed data to a different host\cite{kaplan2020amd}.

\subsection{Architectural Support for Sealing}
Solutions like OP-TEE~\cite{optee}, Keystone\cite{lee2020keystone}, TIMBER-V~\cite{weiser2019timber} and Sanctuary~\cite{brasser2019sanctuary} 
provide sealing support through their software TCB. Here, the TCB exposes an interface to create sealing keys for each \enclave. No additional hardware or architectural support is required in these cases. Hence, this technique is potentially applicable to TEE designs that have a run-time TCB component that is implemented in software.
TPM-based solutions require a platform TPM chip that supports the sealing capability discussed above. Since TPMs are usually off-chip components, and are connected over an unprotected bus, such solutions are not secure against \advBus.
Solutions like Intel SGX~\cite{mckeen2013innovative,anati2013innovative} expose special CPU instructions in hardware to enable sealing. More specifically, they include instructions to generate and access sealing keys based on different types of binding (e.g., developer identity, \enclave{} measurement). %

Finally, in all cases, the architecture must ensure that only the TCB and the owner \enclave{} has access to the sealing key(s). These protections could be implemented through the isolation mechanisms described in \Cref{sec:isolation}.

\section{TCB Discussion}

\begin{table*}
    \centering
    \begin{adjustbox}{width=0.8\linewidth}
    \begin{threeparttable}
    \begin{tabular}{llcccccccrr} \toprule
            & \multirow{2}{*}{Name} & \multirow{2}{*}{RTM} & \multirow{2}{*}{Measurement} & \multirow{2}{*}{Attestation} & \multicolumn{2}{c}{Isolation} & \multirow{2}{*}[-0.5em]{\makecell{Secure \\ IO}} & \multirow{2}{*}{\makecell{Secure \\ storage}} & \multicolumn{2}{c}{TCB size} \\ \cmidrule{6-7} \cmidrule{10-11}
            & & & & & CPU & Memory & & & \multicolumn{1}{c}{Boot} & \multicolumn{1}{c}{Monitor} \\ \midrule
        \multirow{6}{*}{\begin{sideways}Industry\end{sideways}} 
            & Intel SGX~\cite{mckeen2013innovative,anati2013innovative,intelxucode} & I & M & M & M & U+M & - & M & NA & NA \\
            & Intel TDX~\cite{tdxbasespec}     & I & M & M & M & M & - & - & 0 & NA \\ 
            & AMD SEV-SNP\cite{kaplan2020amd}      & I & U & M & U & U & - & U & NA & NA \\
            & ARM TZ\cite{armtz}               & I & M & M & M & M & M & - & 50k LoC & 50k LoC$^\dagger$ \\
            & ARM CCA~\cite{armrealms}         & I & M & M & M & M & M & - & 50k LoC & NA \\
            & IBM PEF~\cite{hunt2021power}     & I & M & M & M & M & - & M & 400k LoC & 75k LoC\\
            \midrule
        \multirow{25}{*}{\begin{sideways}Academia\end{sideways}} 
            & Flicker~\cite{mccune2008flicker}      & I & M & M & M & - & - & I & 0 & 0.25k LoC\\
            & SEA~\cite{mccune2008low}              & I & M & M & M & - & - & I & 0 & NA \\
            & SICE~\cite{azab2011sice}              & I & M & M & M & M & - & - & NA & 2.1k LoC \\
            & PodArch~\cite{shinde2015podarch}      & - & - & - & I & I & - & - & 0 & 0 \\
            & HyperCoffer~\cite{xia2013architecture}& - & - & - & I & I & - & - & NA & 1.1k LoC\\
            & H-SVM~\cite{jin2011architectural,jin2015h} & I & M & M & M & M & - & - & NA & 1.4k LoC \\
            & EqualVisor~\cite{deng2014equalvisor}  & I & - & - & M & M & - & - & NA & 1.2k LoC \\
            & xu-cc15~\cite{xu2015architectural}    & - & - & - & I & I & - & - & 0 & 0\\
            & wen-cf13~\cite{wen2013multi}          & - & - & - & I & I & - & -  & 0 & 0 \\
            & Komodo~\cite{ferraiuolo2017komodo}    & I & M & M & M & M & - & - & 0.8k LoC & 2.7k LoC \\
            & SANCTUARY~\cite{brasser2019sanctuary} & I & M & M & M & M & - & M & 50k LoC & 51.5k LoC$^\dagger$ \\
            & TrustICE~\cite{sun2015trustice}       & - & - & - & M & M & M & - & 50k LoC & 0.28k LoC\\
            & HA-VMSI~\cite{zhu2017ha}              & I & M & M & M & M & - & - & NA & 3.5k LoC\\
            & Sanctum~\cite{costan2016sanctum}      & I & M & M & M & M & - & - & 0.4k LoC & 5k LoC\\
            & TIMBER-V~\cite{weiser2019timber}      & I & M & M & M & M & - & M & NA & 2k LoC \\
            & Keystone~\cite{lee2020keystone}       & I & M & M & M & M & - & M & 16.5k LoC & 10k LoC \\
            & Penglai~\cite{feng2021scalable}       & I & M & M & M & M & - & - & 16.5k LoC & 6.4k LoC \\
            & CURE~\cite{bahami2021cure}            & I & M & M & M & M & M & - & 16.5k LoC & 3k LoC \\
            
            & Iso-X~\cite{evtyushkin2014iso}        & I & I & I & I & I & - & - & 0 & 0\\
            & HyperWall~\cite{szefer2012architectural}  & I & I & I & I & I & - & - & 0 & 0 \\
            & Sancus~\cite{noorman2013sancus,noorman2017sancus} & I & I & I & - & I & - & - & 0 & 0 \\
            & TrustLite~\cite{koeberl2014trustlite} & I & M & M & M & M & M & - & NA & NA \\
            & TyTan~\cite{brasser2015tytan}         & I & M & M & M & M & - & M & NA & NA \\
            
            & XOM~\cite{lie2000xom}                 & - & - & - & I & I & - & - & 0 & 0\\
            & AEGIS~\cite{suh2003aegis}             & - & - & - & I & I & - & - & 0 & 0\\
        \bottomrule
    \end{tabular}
    
    \begin{tablenotes}
        \item [$\dagger$] Using OP-TEE\cite{optee} as a base (around 50k LoC).
    \end{tablenotes}
    \caption{TCB comparison for the surveyed TEEs. All individual components are marked to be either mutable (M), immutable (I), unknown (U), or not supported (-).}
    \label{tab:tcb}
    \end{threeparttable}
    \end{adjustbox}
\end{table*}

This section summarizes the TCB composition of existing TEEs designs. However, it is hard to attribute individual design choices to TCB size or complexity. Besides, although the common metric for TCB size is lines-of-code, accurate numbers are not often available for all TCB components.
Therefore, we discuss whether TEEs implement their different architectural components in a completely \emph{immutable} way or whether they include elements that are \emph{mutable}. When a component is mutable, it can be changed post-manufacture and hence, updated during the lifetime of the computing platform which is important to avoid costly product recalls. Such updates of the TCB (also called TCB recovery~\cite{inteltcbrecovery}) are typically reflected in the attestation report of the platform. Mutable components usually are implemented in software and immutable components are implemented in hardware. We note that in some cases, software can be implemented in an immutable manner (e.g., Boot ROM) and hardware components could be mutable (e.g., $\mu$code in certain CPUs). %

\Cref{tab:tcb} summarizes the components in existing TEE designs that include a mutable element (labelled as M). If an entry in the table labelled as immutable (labelled I), it means that element cannot be changed or updated post manufacture. We use U to denote the cases where the mutability of the implementation is unclear.

\subsubsection*{TCB of Verifiable Launch}
The RTM in all TEEs that support one is immutable as expected. In most cases, this RTM is only used to measure the very first (or next in DRTM) software TCB component in the chain of trust which directly or indirectly (through later components in the chain) measures the actual enclave. Very few designs, namely, HyperWall\cite{szefer2012architectural}, Iso-X\cite{evtyushkin2014iso}, and Sancus\cite{noorman2017sancus} perform the entire enclave measurement in hardware. All other TEEs allow updating the measurement procedure, e.g., to include system parameters such as if hyperthreading is enabled\cite{kaplan2020amd,intelias}. Attestation is typically also performed by a mutable part of the software TCB with the same exceptions as the measurement above. %

\subsubsection*{TCB of Run-Time Isolation}
Many TEEs use a mutable part of the software TCB to implement secure context switches for CPU isolation and manage memory isolation. Exceptions that implement CPU and memory isolation completely in hardware are PodArch\cite{shinde2015podarch}, Hypercoffer\cite{xia2013architecture}, H-SVM\cite{jin2011architectural}, Hyperwall\cite{szefer2012architectural}, Iso-X\cite{evtyushkin2014iso}, XOM\cite{lie2000xom}, Aegis\cite{suh2003aegis} and the works by Xu et. al.\cite{xu2015architectural} and Wen et. al.\cite{wen2013multi}. In some cases, the split of this functionality between mutable and immutable components remains unclear~\cite{costan2016intel,buhren2019insecure}. 

\subsubsection*{TCB of Secure IO}
Most of the studied TEEs do not support trusted IO. Exceptions include TrustICE\cite{sun2015trustice}, CURE\cite{bahami2021cure} and Trustlite\cite{koeberl2014trustlite} and they all rely on mutable software TCB for trusted IO. We note that there are several proposals that focus on enabling trusted IO with some of the TEEs that do not natively support trusted IO as discussed in \Cref{sec:io}.

\subsubsection*{TCB of Secure Storage}
Of the 31 TEE proposals that we studied, 10 proposals discuss sealing support explicitly. Some of these such as Flicker\cite{mccune2008flicker}, SEA\cite{mccune2008low} and IBM PEF\cite{hunt2021power} rely on the TPM for this, while others such as Sanctuary\cite{brasser2019sanctuary}, Timber-V\cite{weiser2019timber}, Keystone\cite{lee2020keystone} and TrustLite\cite{koeberl2014trustlite} implement this feature as part of their software TCB. While Intel SGX provides similar functionality\cite{anati2013innovative}, it is unclear if or what part of it is mutable. 

\subsubsection*{Overall TCB Size}
Usually, most TEE designs and implementations seek to minimize the TCB of the TEE architecture to reduce the risk of it being buggy or vulnerable, and in-theory making them more amendable to formal verification. 
However, in-practice, it is hard to ignore the advantage of being able to update a TCB component if necessary, instead of relying on the TCB being bug-free. Furthermore, as shown in this paper, many TEEs use similar mechanisms overall irrespective of whether they are implemented in a mutable or immutable component. Given this and the fact that existing implementations vary widely in terms of the features they support, we caution against using mutable TCB sizes to compare TEEs. We only include the mutable TCB complexity measured in terms of lines of code obtained from our survey here for completeness.

\section{Related Work}

There have been several studies on TEEs that compare them in terms of the types of security protections that they provide; we summarize these below. We limit the following discussion to survey papers on TEEs and exclude the various papers on a single TEE themselves because the latter set of works are the subject of this paper.

Sabt et. al. are among the first to recognize that there were competing TEE definitions around 2015 and therefore, attempt to arrive at a formal TEE definition. Following that, ARM TrustZone-based TEEs from industry and academia were briefly surveyed using this definition\cite{sabt2015trusted}. 
A more detailed discussion of ARM TrustZone-based TEEs, their various flavors across different ARM processor versions, and software solutions that leverage ARM TrustZone can be found in~\cite{pinto2019demystifying}. A further study focuses on the security limitations of ARM TrustZone-based~\cite{cerdeira2020sok}. The primary focus of all these works is on TEE architectures, systems and attacks involving ARM Trustzone. We note that~\cite{pinto2019demystifying} briefly covers a few other non-ARM TEEs but focuses only on the security properties that they enable and not their architectural details. 

There have been similar works that survey security mechanisms and TEEs in the RISC-V ecosystem. In~\cite{lu2021survey}, the author surveys hardware and architectural security for RISC-V processors and  contrasts them to ARM processors. While this paper makes many good comparisons between the ARM and RISC-V architectures in general (e.g., exception levels) as well as with respect to security-related features (e.g., support for cryptography, ISA extensions), it only mentions that the Keystone architecture\cite{lee2020keystone} is similar to ARM TrustZone but defers any further discussion to future work. More recently, existing RISC-V TEE architectures are summarized in~\cite{dessoukyenclave} but no comparison to TEEs on different processor architectures is given. Similar limitations apply to previous studies on Intel SGX and its applications~\cite{zheng2021survey} as well as security limitations~\cite{fei2021security, randmets2021overview}. In contrast to these surveys that focus on ARM or RISC-V, this paper covers TEEs across various various instruction set architectures and compares the different micro-architectural elements that underpin them. 

The only previous effort at systematization of TEE architectures among all major processor architectures is~\cite{zhao2019sok}. Here, the authors summarize TEEs architectures on four dimensions: mechanisms to ensure integrity of the initial contents of the TEE, memory protection, scope (e.g., per system, processor package, core or thread) and finally, developer access. It discusses the security implications of incomplete hardware abstractions that do not capture implementation aspects of TEEs (e.g., timing of operations, caching, concurrency) and finally, covers different applications of TEEs. The discussion in~\cite{zhao2019sok} of micro-architectural support for run-time isolation mechanisms in TEEs is limited to memory protection, and, for example, does not include CPU-state protection or trusted IO support. Furthermore, the discussion on memory protection is also not exhaustive and as detailed as covered in this paper. 

Existing literature also includes surveys of TEE architectures (e.g.,~\cite{coppolino2019comprehensive,jauernig2020trusted}) as well as security properties (e.g., ~\cite{demigha2021hardware, valadares2021trusted}). Although these works cover TEEs across multiple processor architectures, none of them systematically catalog and analyze the various architectural design decisions underlying these TEE designs and analyze their advantages and disadvantages as we do in this paper.

\section{Conclusion}

In this paper, we analyzed the underlying design choices of commercial and academic TEEs for four high-level security goals: (i) verifiable launch, (ii) run-time isolation, (iii) secure IO, and (iv) secure storage. Even though these proposals often seem very different at first, many of them share many design decisions but use different names and descriptions for them. We believe our findings can help upcoming TEE proposals weigh the different design decisions and hopefully reduces reinvention in the field.
\bibliographystyle{IEEEtran}
\bibliography{IEEEabrv,bibliography/surveys,bibliography/tees,bibliography/attacks,bibliography/defenses, bibliography/trustedio,bibliography/other}

\begin{thebibliography}{10}
\providecommand{\url}[1]{#1}
\csname url@samestyle\endcsname
\providecommand{\newblock}{\relax}
\providecommand{\bibinfo}[2]{#2}
\providecommand{\BIBentrySTDinterwordspacing}{\spaceskip=0pt\relax}
\providecommand{\BIBentryALTinterwordstretchfactor}{4}
\providecommand{\BIBentryALTinterwordspacing}{\spaceskip=\fontdimen2\font plus
\BIBentryALTinterwordstretchfactor\fontdimen3\font minus
  \fontdimen4\font\relax}
\providecommand{\BIBforeignlanguage}[2]{{%
\expandafter\ifx\csname l@#1\endcsname\relax
\typeout{** WARNING: IEEEtran.bst: No hyphenation pattern has been}%
\typeout{** loaded for the language `#1'. Using the pattern for}%
\typeout{** the default language instead.}%
\else
\language=\csname l@#1\endcsname
\fi
#2}}
\providecommand{\BIBdecl}{\relax}
\BIBdecl

\bibitem{tpm}
T.~C. Group, ``{TCG} specification architecture overview,'' Trusted Computing
  Group, Tech. Rep., Aug. 2007, revision 1.4.

\bibitem{costan2016intel}
V.~Costan and S.~Devadas, ``Intel {SGX} explained.'' \emph{IACR Cryptol. ePrint
  Arch.}, vol. 2016, no.~86, pp. 1--118, 2016.

\bibitem{kaplan2020amd}
D.~Kaplan, J.~Powell, and T.~Woller, ``{AMD SEV-SNP}: Strengthening {VM}
  isolation with integrity protection and more,'' AMD, Tech. Rep., 2020.

\bibitem{costan2016sanctum}
V.~Costan, I.~Lebedev, and S.~Devadas, ``Sanctum: Minimal hardware extensions
  for strong software isolation,'' in \emph{25th $\{$USENIX$\}$ Security
  Symposium ($\{$USENIX$\}$ Security 16)}, 2016, pp. 857--874.

\bibitem{mccune2008flicker}
J.~M. McCune, B.~J. Parno, A.~Perrig, M.~K. Reiter, and H.~Isozaki, ``Flicker:
  An execution infrastructure for {TCB} minimization,'' in \emph{Proceedings of
  the 3rd ACM SIGOPS/EuroSys European Conference on Computer Systems 2008},
  2008, pp. 315--328.

\bibitem{mccune2008low}
J.~M. McCune, B.~Parno, A.~Perrig, M.~K. Reiter, and A.~Seshadri, ``How low can
  you go? recommendations for hardware-supported minimal {TCB} code
  execution,'' \emph{ACM SIGOPS Operating Systems Review}, vol.~42, no.~2, pp.
  14--25, 2008.

\bibitem{mckeen2013innovative}
F.~McKeen, I.~Alexandrovich, A.~Berenzon, C.~V. Rozas, H.~Shafi, V.~Shanbhogue,
  and U.~R. Savagaonkar, ``Innovative instructions and software model for
  isolated execution.'' in \emph{Hardware and Architectural Support for
  Security and Privacy}, 2013, pp. 1--8.

\bibitem{tdxbasespec}
I.~Corporation, ``{Intel®} trust domain extensions {(Intel® TDX)} module base
  architecture specification,'' Intel Corporation, Tech. Rep., Sep. 2021,
  348549-001US.

\bibitem{armrealms}
Arm, ``{Arm®} architecture reference manual supplement, the realm management
  extension {(RME)}, for {Armv9-A},'' document number: {ARM DDI} 0615.

\bibitem{lee2020off}
D.~Lee, D.~Jung, I.~T. Fang, C.-C. Tsai, and R.~A. Popa, ``An off-chip attack
  on hardware enclaves via the memory bus,'' in \emph{29th {USENIX} Security
  Symposium ({USENIX} Security 20)}, 2020.

\bibitem{krachenfels2021laser}
T.~Krachenfels, T.~Kiyan, S.~Tajik, and J.-P. Seifert, ``Automatic extraction
  of secrets from the transistor jungle using laser-assisted side-channel
  attacks,'' in \emph{30th {USENIX} Security Symposium ({USENIX} Security
  21)}.\hskip 1em plus 0.5em minus 0.4em\relax {USENIX} Association, Aug. 2021,
  pp. 627--644.

\bibitem{kocher1999differential}
P.~Kocher, J.~Jaffe, and B.~Jun, ``Differential power analysis,'' in
  \emph{Annual international cryptology conference}.\hskip 1em plus 0.5em minus
  0.4em\relax Springer, 1999, pp. 388--397.

\bibitem{liu2015last}
F.~Liu, Y.~Yarom, Q.~Ge, G.~Heiser, and R.~B. Lee, ``Last-level cache
  side-channel attacks are practical,'' in \emph{2015 IEEE symposium on
  security and privacy}.\hskip 1em plus 0.5em minus 0.4em\relax IEEE, 2015, pp.
  605--622.

\bibitem{brasser2017software}
F.~Brasser, U.~M{\"u}ller, A.~Dmitrienko, K.~Kostiainen, S.~Capkun, and A.-R.
  Sadeghi, ``Software grand exposure: {SGX} cache attacks are practical,'' in
  \emph{11th {USENIX} Workshop on Offensive Technologies ({WOOT} 17)}, 2017.

\bibitem{li2021cipherleaks}
M.~Li, Y.~Zhang, H.~Wang, K.~Li, and Y.~Cheng, ``{CIPHERLEAKS}: Breaking
  constant-time cryptography on {AMD} {SEV} via the ciphertext side channel,''
  in \emph{30th USENIX Security Symposium (USENIX Security 21)}.\hskip 1em plus
  0.5em minus 0.4em\relax USENIX Association, Aug. 2021, pp. 717--732.

\bibitem{ahmad2019obfuscuro}
A.~Ahmad, B.~Joe, Y.~Xiao, Y.~Zhang, I.~Shin, and B.~Lee, ``Obfuscuro: A
  commodity obfuscation engine on {Intel} {SGX},'' in \emph{Network and
  Distributed System Security Symposium}, 2019.

\bibitem{rane2015raccoon}
A.~Rane, C.~Lin, and M.~Tiwari, ``Raccoon: Closing digital side-channels
  through obfuscated execution,'' in \emph{24th $\{$USENIX$\}$ Security
  Symposium ($\{$USENIX$\}$ Security 15)}, 2015, pp. 431--446.

\bibitem{deng2014equalvisor}
L.~Deng, Q.~Zeng, W.~Wang, and Y.~Liu, ``{EqualVisor}: Providing memory
  protection in an untrusted commodity hypervisor,'' in \emph{2014 IEEE 13th
  International Conference on Trust, Security and Privacy in Computing and
  Communications}.\hskip 1em plus 0.5em minus 0.4em\relax IEEE, 2014, pp.
  300--309.

\bibitem{kauer2007oslo}
B.~Kauer, ``{OSLO}: Improving the security of trusted computing.'' in
  \emph{USENIX Security Symposium}, vol.~24, 2007, p. 173.

\bibitem{anati2013innovative}
I.~Anati, S.~Gueron, S.~Johnson, and V.~Scarlata, ``Innovative technology for
  cpu based attestation and sealing,'' in \emph{Proceedings of the 2nd
  international workshop on hardware and architectural support for security and
  privacy}, vol.~13.\hskip 1em plus 0.5em minus 0.4em\relax Citeseer, 2013,
  p.~7.

\bibitem{armtz}
T.~Alves and D.~Felton, ``{TrustZone}: Integrated hardware and software
  security - enabling trusted computing in embedded systems,'' \emph{White
  paper}, 2004.

\bibitem{hunt2021power}
G.~D. Hunt, R.~Pai, M.~V. Le, H.~Jamjoom, S.~Bhattiprolu, R.~Boivie, L.~Dufour,
  B.~Frey, M.~Kapur, K.~A. Goldman \emph{et~al.}, ``Confidential computing for
  {OpenPOWER},'' in \emph{Proceedings of the Sixteenth European Conference on
  Computer Systems}, 2021, pp. 294--310.

\bibitem{azab2011sice}
A.~M. Azab, P.~Ning, and X.~Zhang, ``{SICE}: a hardware-level strongly isolated
  computing environment for x86 multi-core platforms,'' in \emph{Proceedings of
  the 18th ACM conference on Computer and communications security}, 2011, pp.
  375--388.

\bibitem{shinde2015podarch}
S.~Shinde, S.~Tople, D.~Kathayat, and P.~Saxena, ``Podarch: Protecting legacy
  applications with a purely hardware {TCB},'' \emph{National University of
  Singapore, Tech. Rep}, 2015.

\bibitem{xia2013architecture}
Y.~Xia, Y.~Liu, and H.~Chen, ``Architecture support for guest-transparent vm
  protection from untrusted hypervisor and physical attacks,'' in \emph{2013
  IEEE 19th International Symposium on High Performance Computer Architecture
  (HPCA)}.\hskip 1em plus 0.5em minus 0.4em\relax IEEE, 2013, pp. 246--257.

\bibitem{jin2011architectural}
S.~Jin, J.~Ahn, S.~Cha, and J.~Huh, ``Architectural support for secure
  virtualization under a vulnerable hypervisor,'' in \emph{2011 44th Annual
  IEEE/ACM International Symposium on Microarchitecture (MICRO)}.\hskip 1em
  plus 0.5em minus 0.4em\relax IEEE, 2011, pp. 272--283.

\bibitem{jin2015h}
S.~Jin, J.~Ahn, J.~Seol, S.~Cha, J.~Huh, and S.~Maeng, ``{H-SVM}:
  Hardware-assisted secure virtual machines under a vulnerable hypervisor,''
  \emph{IEEE Transactions on Computers}, vol.~64, no.~10, pp. 2833--2846, 2015.

\bibitem{xu2015architectural}
L.~Xu, J.~Lee, S.~H. Kim, Q.~Zheng, S.~Xu, T.~Suh, W.~W. Ro, and W.~Shi,
  ``Architectural protection of application privacy against software and
  physical attacks in untrusted cloud environment,'' \emph{IEEE Transactions on
  Cloud Computing}, vol.~6, no.~2, pp. 478--491, 2015.

\bibitem{wen2013multi}
Y.~Wen, J.~Lee, Z.~Liu, Q.~Zheng, W.~Shi, S.~Xu, and T.~Suh, ``Multi-processor
  architectural support for protecting virtual machine privacy in untrusted
  cloud environment,'' in \emph{Proceedings of the ACM International Conference
  on Computing Frontiers}, 2013, pp. 1--10.

\bibitem{ferraiuolo2017komodo}
A.~Ferraiuolo, A.~Baumann, C.~Hawblitzel, and B.~Parno, ``Komodo: Using
  verification to disentangle secure-enclave hardware from software,'' in
  \emph{Proceedings of the 26th Symposium on Operating Systems Principles},
  2017, pp. 287--305.

\bibitem{brasser2019sanctuary}
F.~Brasser, D.~Gens, P.~Jauernig, A.-R. Sadeghi, and E.~Stapf, ``{SANCTUARY}:
  {ARMing} {TrustZone} with user-space enclaves.'' in \emph{NDSS}, 2019.

\bibitem{sun2015trustice}
H.~Sun, K.~Sun, Y.~Wang, J.~Jing, and H.~Wang, ``{TrustICE}: Hardware-assisted
  isolated computing environments on mobile devices,'' in \emph{2015 45th
  Annual IEEE/IFIP International Conference on Dependable Systems and
  Networks}.\hskip 1em plus 0.5em minus 0.4em\relax IEEE, 2015, pp. 367--378.

\bibitem{zhu2017ha}
M.~Zhu, B.~Tu, W.~Wei, and D.~Meng, ``{HA-VMSI}: A lightweight virtual machine
  isolation approach with commodity hardware for {ARM},'' \emph{ACM SIGPLAN
  Notices}, vol.~52, no.~7, pp. 242--256, 2017.

\bibitem{weiser2019timber}
S.~Weiser, M.~Werner, F.~Brasser, M.~Malenko, S.~Mangard, and A.-R. Sadeghi,
  ``{TIMBER-V}: Tag-isolated memory bringing fine-grained enclaves to
  {RISC-V}.'' in \emph{NDSS}, 2019.

\bibitem{lee2020keystone}
D.~Lee, D.~Kohlbrenner, S.~Shinde, K.~Asanovi{\'c}, and D.~Song, ``Keystone: An
  open framework for architecting trusted execution environments,'' in
  \emph{Proceedings of the Fifteenth European Conference on Computer Systems},
  2020, pp. 1--16.

\bibitem{feng2021scalable}
E.~Feng, X.~Lu, D.~Du, B.~Yang, X.~Jiang, Y.~Xia, B.~Zang, and H.~Chen,
  ``Scalable memory protection in the {PENGLAI} enclave,'' in \emph{15th
  {USENIX} Symposium on Operating Systems Design and Implementation ({OSDI}
  21)}, 2021, pp. 275--294.

\bibitem{bahami2021cure}
R.~Bahmani, F.~Brasser, G.~Dessouky, P.~Jauernig, M.~Klimmek, A.-R. Sadeghi,
  and E.~Stapf, ``{CURE}: A security architecture with customizable and
  resilient enclaves,'' in \emph{30th {USENIX} Security Symposium ({USENIX}
  Security 21)}.\hskip 1em plus 0.5em minus 0.4em\relax {USENIX} Association,
  Aug. 2021, pp. 1073--1090.

\bibitem{evtyushkin2014iso}
D.~Evtyushkin, J.~Elwell, M.~Ozsoy, D.~Ponomarev, N.~A. Ghazaleh, and R.~Riley,
  ``{Iso-X}: A flexible architecture for hardware-managed isolated execution,''
  in \emph{2014 47th Annual IEEE/ACM International Symposium on
  Microarchitecture}.\hskip 1em plus 0.5em minus 0.4em\relax IEEE, 2014, pp.
  190--202.

\bibitem{szefer2012architectural}
J.~Szefer and R.~B. Lee, ``Architectural support for hypervisor-secure
  virtualization,'' \emph{ACM SIGPLAN Notices}, vol.~47, no.~4, pp. 437--450,
  2012.

\bibitem{noorman2013sancus}
J.~Noorman, P.~Agten, W.~Daniels, R.~Strackx, A.~Van~Herrewege, C.~Huygens,
  B.~Preneel, I.~Verbauwhede, and F.~Piessens, ``Sancus: Low-cost trustworthy
  extensible networked devices with a zero-software trusted computing base,''
  in \emph{22nd {USENIX} Security Symposium ({USENIX} Security 13)}, 2013, pp.
  479--498.

\bibitem{noorman2017sancus}
J.~Noorman, J.~V. Bulck, J.~T. M{\"u}hlberg, F.~Piessens, P.~Maene, B.~Preneel,
  I.~Verbauwhede, J.~G{\"o}tzfried, T.~M{\"u}ller, and F.~Freiling, ``Sancus
  2.0: A low-cost security architecture for {IoT} devices,'' \emph{ACM
  Transactions on Privacy and Security (TOPS)}, vol.~20, no.~3, pp. 1--33,
  2017.

\bibitem{koeberl2014trustlite}
P.~Koeberl, S.~Schulz, A.-R. Sadeghi, and V.~Varadharajan, ``Trustlite: A
  security architecture for tiny embedded devices,'' in \emph{Proceedings of
  the Ninth European Conference on Computer Systems}, 2014, pp. 1--14.

\bibitem{brasser2015tytan}
F.~Brasser, B.~El~Mahjoub, A.-R. Sadeghi, C.~Wachsmann, and P.~Koeberl,
  ``{TyTAN}: Tiny trust anchor for tiny devices,'' in \emph{Proceedings of the
  52nd annual design automation conference}, 2015, pp. 1--6.

\bibitem{lie2000xom}
D.~Lie, C.~Thekkath, M.~Mitchell, P.~Lincoln, D.~Boneh, J.~Mitchell, and
  M.~Horowitz, ``Architectural support for copy and tamper resistant
  software,'' \emph{Acm Sigplan Notices}, vol.~35, no.~11, pp. 168--177, 2000.

\bibitem{suh2003aegis}
G.~E. Suh, D.~Clarke, B.~Gassend, M.~Van~Dijk, and S.~Devadas, ``Aegis:
  Architecture for tamper-evident and tamper-resistant processing,'' in
  \emph{ACM International Conference on Supercomputing 25th Anniversary
  Volume}, 2003, pp. 357--368.

\bibitem{lebedev2018sanctumattest}
I.~Lebedev, K.~Hogan, and S.~Devadas, ``Secure boot and remote attestation in
  the {Sanctum} processor,'' in \emph{2018 IEEE 31st Computer Security
  Foundations Symposium (CSF)}.\hskip 1em plus 0.5em minus 0.4em\relax IEEE,
  2018, pp. 46--60.

\bibitem{suh2003efficient}
G.~E. Suh, D.~Clarke, B.~Gasend, M.~Van~Dijk, and S.~Devadas, ``Efficient
  memory integrity verification and encryption for secure processors,'' in
  \emph{Proceedings. 36th Annual IEEE/ACM International Symposium on
  Microarchitecture, 2003. MICRO-36.}\hskip 1em plus 0.5em minus 0.4em\relax
  IEEE, 2003, pp. 339--350.

\bibitem{bourgeat2019mi6}
T.~Bourgeat, I.~Lebedev, A.~Wright, S.~Zhang, and S.~Devadas, ``Mi6: Secure
  enclaves in a speculative out-of-order processor,'' in \emph{Proceedings of
  the 52nd Annual IEEE/ACM International Symposium on Microarchitecture}, 2019,
  pp. 42--56.

\bibitem{li2018vbutton}
W.~Li, S.~Luo, Z.~Sun, Y.~Xia, L.~Lu, H.~Chen, B.~Zang, and H.~Guan,
  ``{VButton}: Practical attestation of user-driven operations in mobile
  apps,'' in \emph{Proceedings of the 16th annual international conference on
  mobile systems, applications, and services}, 2018, pp. 28--40.

\bibitem{ying2018truz}
K.~Ying, A.~Ahlawat, B.~Alsharifi, Y.~Jiang, P.~Thavai, and W.~Du,
  ``{TruZ-Droid}: Integrating {TrustZone} with mobile operating system,'' in
  \emph{Proceedings of the 16th annual international conference on mobile
  systems, applications, and services}, 2018, pp. 14--27.

\bibitem{nasahl2021hector}
P.~Nasahl, R.~Schilling, M.~Werner, and S.~Mangard, ``{HECTOR-V}: A
  heterogeneous {CPU} architecture for a secure {RISC-V} execution
  environment,'' in \emph{Proceedings of the 2021 ACM Asia Conference on
  Computer and Communications Security}, 2021, pp. 187--199.

\bibitem{zhou2012building}
Z.~Zhou, V.~D. Gligor, J.~Newsome, and J.~M. McCune, ``Building verifiable
  trusted path on commodity x86 computers,'' in \emph{2012 IEEE symposium on
  security and privacy}.\hskip 1em plus 0.5em minus 0.4em\relax IEEE, 2012, pp.
  616--630.

\bibitem{li2015adattester}
W.~Li, H.~Li, H.~Chen, and Y.~Xia, ``{AdAttester}: Secure online mobile
  advertisement attestation using {TrustZone},'' in \emph{Proceedings of the
  13th annual international conference on mobile systems, applications, and
  services}, 2015, pp. 75--88.

\bibitem{li2014building}
W.~Li, M.~Ma, J.~Han, Y.~Xia, B.~Zang, C.-K. Chu, and T.~Li, ``Building trusted
  path on untrusted device drivers for mobile devices,'' in \emph{Proceedings
  of 5th Asia-Pacific Workshop on Systems}, 2014, pp. 1--7.

\bibitem{sun2015trustotp}
H.~Sun, K.~Sun, Y.~Wang, and J.~Jing, ``{TrustOTP}: Transforming smartphones
  into secure one-time password tokens,'' in \emph{Proceedings of the 22nd ACM
  SIGSAC Conference on Computer and Communications Security}, 2015, pp.
  976--988.

\bibitem{jang2019heterogeneous}
I.~Jang, A.~Tang, T.~Kim, S.~Sethumadhavan, and J.~Huh, ``Heterogeneous
  isolated execution for commodity {GPUs},'' in \emph{Proceedings of the
  Twenty-Fourth International Conference on Architectural Support for
  Programming Languages and Operating Systems}, 2019, pp. 455--468.

\bibitem{eskandarian2019fidelius}
S.~Eskandarian, J.~Cogan, S.~Birnbaum, P.~C.~W. Brandon, D.~Franke, F.~Fraser,
  G.~Garcia, E.~Gong, H.~T. Nguyen, T.~K. Sethi \emph{et~al.}, ``Fidelius:
  Protecting user secrets from compromised browsers,'' in \emph{2019 IEEE
  Symposium on Security and Privacy (SP)}.\hskip 1em plus 0.5em minus
  0.4em\relax IEEE, 2019, pp. 264--280.

\bibitem{peters2018bastion}
T.~Peters, R.~Lal, S.~Varadarajan, P.~Pappachan, and D.~Kotz, ``{BASTION-SGX}:
  Bluetooth and architectural support for trusted {I/O} on {SGX},'' in
  \emph{Proceedings of the 7th International Workshop on Hardware and
  Architectural Support for Security and Privacy}, 2018, pp. 1--9.

\bibitem{dhar2019protection}
\BIBentryALTinterwordspacing
A.~Dhar, E.~Ulqinaku, K.~Kostiainen, and S.~Capkun, ``{ProtectIOn}:
  Root-of-trust for {IO} in compromised platforms,'' in \emph{27th Annual
  Network and Distributed System Security Symposium, {NDSS} 2020, San Diego,
  California, USA, February 23-26, 2020}.\hskip 1em plus 0.5em minus
  0.4em\relax The Internet Society, 2020. [Online]. Available:
  \url{https://www.ndss-symposium.org/ndss-paper/protection-root-of-trust-for-io-in-compromised-platforms/}
\BIBentrySTDinterwordspacing

\bibitem{zhu2019enabling}
J.~Zhu, R.~Hou, X.~Wang, W.~Wang, J.~Cao, L.~Zhao, F.~Yuan, P.~Li, Z.~Wang,
  B.~Zhao \emph{et~al.}, ``Enabling privacy-preserving, compute-and
  data-intensive computing using heterogeneous trusted execution environment,''
  \emph{arXiv preprint arXiv:1904.04782}, 2019.

\bibitem{synopsysPCIencryption}
\BIBentryALTinterwordspacing
Synopsys, ``Security modules for standard interfaces,'' accessed on 2021-11-29.
  [Online]. Available:
  \url{https://www.synopsys.com/designware-ip/security-ip/interface-security-modules.html}
\BIBentrySTDinterwordspacing

\bibitem{weiser2017sgxio}
S.~Weiser and M.~Werner, ``{SGXIO}: Generic trusted {I/O} path for {Intel
  SGX},'' in \emph{Proceedings of the seventh ACM on conference on data and
  application security and privacy}, 2017, pp. 261--268.

\bibitem{kwon2019zerokernel}
O.~Kwon, Y.~Kim, J.~Huh, and H.~Yoon, ``{ZeroKernel}: Secure context-isolated
  execution on commodity {GPUs},'' \emph{IEEE Transactions on Dependable and
  Secure Computing}, 2019.

\bibitem{oh2021meetgo}
H.~Oh, K.~Nam, S.~Jeon, Y.~Cho, and Y.~Paek, ``{MeetGo}: A trusted execution
  environment for remote applications on {FPGA},'' \emph{IEEE Access}, vol.~9,
  pp. 51\,313--51\,324, 2021.

\bibitem{zhao2021shef}
M.~Zhao, M.~Gao, and C.~Kozyrakis, ``{ShEF}: Shielded enclaves for cloud
  {FPGAs},'' \emph{arXiv preprint arXiv:2103.03500}, 2021.

\bibitem{volos2018graviton}
S.~Volos, K.~Vaswani, and R.~Bruno, ``Graviton: Trusted execution environments
  on {GPUs},'' in \emph{13th {USENIX} Symposium on Operating Systems Design and
  Implementation ({OSDI} 18)}, 2018, pp. 681--696.

\bibitem{hunt2020telekine}
T.~Hunt, Z.~Jia, V.~Miller, A.~Szekely, Y.~Hu, C.~J. Rossbach, and E.~Witchel,
  ``Telekine: Secure computing with cloud {GPUs},'' in \emph{17th
  $\{$USENIX$\}$ Symposium on Networked Systems Design and Implementation
  ($\{$NSDI$\}$ 20)}, 2020, pp. 817--833.

\bibitem{wang2020segive}
Z.~Wang, F.~Zheng, J.~Lin, G.~Fan, and J.~Dong, ``{SEGIVE}: A practical
  framework of secure {GPU} execution in virtualization environment,'' in
  \emph{2020 IEEE 39th International Performance Computing and Communications
  Conference (IPCCC)}.\hskip 1em plus 0.5em minus 0.4em\relax IEEE, 2020, pp.
  1--10.

\bibitem{oh2020trustore}
H.~Oh, A.~Ahmad, S.~Park, B.~Lee, and Y.~Paek, ``{TrustOre}: Side-channel
  resistant storage for {SGX} using {Intel} hybrid {CPU-FGPA},'' in
  \emph{Proceedings of the 2020 ACM SIGSAC Conference on Computer and
  Communications Security}, 2020, pp. 1903--1918.

\bibitem{epidIntel2016}
S.~Johnson, V.~Scarlata, C.~Rozas, E.~Brickell, and F.~Mckeen, ``Intel software
  guard extensions: {EPID} provisioning and attestation services,'' Intel
  Corperation, Tech. Rep., 2016.

\bibitem{optee}
TrustedFirmware, ``{OP-TEE} documentation,''
  https://optee.readthedocs.io/en/latest/index.html, accessed: 11.11.2021.

\bibitem{intelxucode}
\BIBentryALTinterwordspacing
Intel, ``Xucode: An innovative technology for implementing complex instruction
  flows.'' [Online]. Available:
  \url{https://www.intel.com/content/www/us/en/developer/articles/technical/software-security-guidance/secure-coding/xucode-implementing-complex-instruction-flows.html}
\BIBentrySTDinterwordspacing

\bibitem{inteltcbrecovery}
------, ``Intel® {SGX} trusted computing base ({TCB}) recovery,'' 2018.

\bibitem{intelias}
------, ``Attestation service for intel® software guard extensions ({Intel®
  SGX}): {API} documentation,''
  https://api.trustedservices.intel.com/documents/IAS-API-Spec-rev-4.0.pdf,
  revision 4.1.

\bibitem{buhren2019insecure}
R.~Buhren, C.~Werling, and J.-P. Seifert, ``Insecure until proven updated:
  analyzing {AMD SEV}'s remote attestation,'' in \emph{Proceedings of the 2019
  ACM SIGSAC Conference on Computer and Communications Security}, 2019, pp.
  1087--1099.

\bibitem{sabt2015trusted}
M.~Sabt, M.~Achemlal, and A.~Bouabdallah, ``Trusted execution environment: what
  it is, and what it is not,'' in \emph{2015 IEEE Trustcom/BigDataSE/ISPA},
  vol.~1.\hskip 1em plus 0.5em minus 0.4em\relax IEEE, 2015, pp. 57--64.

\bibitem{pinto2019demystifying}
S.~Pinto and N.~Santos, ``Demystifying {Arm} {TrustZone}: A comprehensive
  survey,'' \emph{ACM Computing Surveys (CSUR)}, vol.~51, no.~6, pp. 1--36,
  2019.

\bibitem{cerdeira2020sok}
D.~Cerdeira, N.~Santos, P.~Fonseca, and S.~Pinto, ``{SoK}: Understanding the
  prevailing security vulnerabilities in {TrustZone}-assisted {TEE} systems,''
  in \emph{2020 IEEE Symposium on Security and Privacy (SP)}.\hskip 1em plus
  0.5em minus 0.4em\relax IEEE, 2020, pp. 1416--1432.

\bibitem{lu2021survey}
T.~Lu, ``A survey on {RISC-V} security: Hardware and architecture,''
  \emph{arXiv preprint arXiv:2107.04175}, 2021.

\bibitem{dessoukyenclave}
G.~Dessouky, A.-R. Sadeghi, and E.~Stapf, ``Enclave computing on {RISC-V}: A
  brighter future for security?'' in \emph{Workshop on Secure {RISC-V}
  Architecture Design {(SECRISC-V'20)}}, Aug 2020.

\bibitem{zheng2021survey}
W.~Zheng, Y.~Wu, X.~Wu, C.~Feng, Y.~Sui, X.~Luo, and Y.~Zhou, ``A survey of
  {Intel SGX} and its applications,'' \emph{Frontiers of Computer Science},
  vol.~15, no.~3, pp. 1--15, 2021.

\bibitem{fei2021security}
S.~Fei, Z.~Yan, W.~Ding, and H.~Xie, ``Security vulnerabilities of {SGX} and
  countermeasures: A survey,'' \emph{ACM Computing Surveys (CSUR)}, vol.~54,
  no.~6, pp. 1--36, 2021.

\bibitem{randmets2021overview}
J.~Randmets, ``An overview of vulnerabilities and mitigations of {Intel SGX}
  applications,'' Cybernetica AS, Tech. Rep., 2021.

\bibitem{zhao2019sok}
L.~Zhao, H.~Shuang, S.~Xu, W.~Huang, R.~Cui, P.~Bettadpur, and D.~Lie, ``{SoK}:
  Hardware security support for trustworthy execution,'' \emph{arXiv preprint
  arXiv:1910.04957}, 2019.

\bibitem{coppolino2019comprehensive}
L.~Coppolino, S.~D’Antonio, G.~Mazzeo, and L.~Romano, ``A comprehensive
  survey of hardware-assisted security: From the edge to the cloud,''
  \emph{Internet of Things}, vol.~6, p. 100055, 2019.

\bibitem{jauernig2020trusted}
P.~Jauernig, A.-R. Sadeghi, and E.~Stapf, ``Trusted execution environments:
  properties, applications, and challenges,'' \emph{IEEE Security \& Privacy},
  vol.~18, no.~2, pp. 56--60, 2020.

\bibitem{demigha2021hardware}
O.~Demigha and R.~Larguet, ``Hardware-based solutions for trusted cloud
  computing,'' \emph{Computers \& Security}, p. 102117, 2021.

\bibitem{valadares2021trusted}
D.~C.~G. Valadares, N.~C. Will, M.~A. Spohn, D.~F. de~Souza~Santos,
  A.~Perkusich, and K.~C. Gorgonio, ``Trusted execution environments for
  cloud/fog-based internet of things applications.'' in \emph{CLOSER}, 2021,
  pp. 111--121.

\bibitem{tdxwhitepaper}
I.~Corporation, ``Intel® trust domain extensions,'' Intel Corperation, Tech.
  Rep., Aug. 2020, 343961-002US.

\bibitem{amdsevapi}
AMD, ``Secure encrypted virtualization {API} version 0.24,'' AMD, Tech. Rep.,
  2020, issue Date: April, 2020.

\bibitem{muhlberg2015lightweight}
J.~T. M{\"u}hlberg, J.~Noorman, and F.~Piessens, ``Lightweight and flexible
  trust assessment modules for the internet of things,'' in \emph{European
  Symposium on Research in Computer Security}.\hskip 1em plus 0.5em minus
  0.4em\relax Springer, 2015, pp. 503--520.

\bibitem{riscv2019privspec}
A.~Waterman, Y.~Lee, R.~Avizienis, D.~A. Patterson, and K.~Asanovi{\'c}, ``The
  {RISC-V} instruction set manual volume {II}: Privileged architecture,''
  \emph{EECS Department, University of California, Berkeley}, 2019.

\bibitem{intelSDM}
\BIBentryALTinterwordspacing
{Intel Corporation}, ``Intel 64 and {IA-32} architectures software developer
  manuals,'' Jun 2021. [Online]. Available:
  \url{https://software.intel.com/content/www/us/en/develop/articles/intel-sdm.html}
\BIBentrySTDinterwordspacing

\bibitem{amdIsa}
\BIBentryALTinterwordspacing
AMD, ``{AMD64} architecture programmer’s manual: Volumes 1-5,'' revision
  4.03. [Online]. Available:
  \url{https://www.amd.com/system/files/TechDocs/40332.pdf}
\BIBentrySTDinterwordspacing

\bibitem{armv8}
``{Arm}® architecture reference manual: {Armv8}, for {Armv8-A} architecture
  profile.''

\end{thebibliography}

\appendix
\section{Architectural Details}
In this section we highlight some implementation details of the surveyed TEEs. For the full details of any given architecture, we refer the reader to the primary references for the specific architectures.

\begin{table*}[tbp]
    \centering
    \begin{threeparttable}
    \begin{tabular}{llcccccclccc} \toprule
            & \multirow{2}{*}[-5pt]{Name} & \multicolumn{6}{c}{Adversary} && \multicolumn{3}{c}{Trusted Access Control Information} \\ \cmidrule{3-8} \cmidrule{10-12}
            & & \advApp{} & \advSSW{} & \advTEE{} & \advBoot{} & \advPer  & \advBus && MPU & Page Tables & Extra Metadata \\ \midrule
        \multirow{6}{*}{\begin{sideways}Industry\end{sideways}} 
            & Intel SGX~\cite{mckeen2013innovative,anati2013innovative} & SL & SL & STL & SL & STL & C && \yes & \no & \yes \\
            & Intel TDX~\cite{tdxbasespec}   & SL & SL & STL & SL & STL & C$^\dagger$ && \yes & \yes & \yes \\ 
            & AMD SEV-SNP\cite{kaplan2020amd}  & STC & STC & STC & - & STL &  C$^\dagger$ && \no & \no & \yes \\
            & ARM TZ\cite{armtz}             & SL & SL & STL & - & SL &  C$^\ddagger$ && \yes & \yes & \no \\
            & ARM CCA~\cite{armrealms}    & STL & STL & STL & - & SL &  C$^\ddagger$ && \no & \yes & \yes \\ 
            & IBM PEF~\cite{hunt2021power}     & SL & SL & STL & - & -  & - && \yes & \yes & \no \\  
            \midrule
        \multirow{25}{*}{\begin{sideways}Academia\end{sideways}} 
            & Flicker~\cite{mccune2008flicker}   & TL & TL & - & SL & SL & - && \no & \no & \no \\
            & SEA~\cite{mccune2008low}           & TL & TL & - & SL & SL & - && \no & \no & \no \\
            & SICE~\cite{azab2011sice}           & SL & SL & SL & - & SL & - && \yes & \no & \no \\
            & PodArch~\cite{shinde2015podarch}   & STL & STL & STL & - & SC & - && \no & \yes & \yes \\
            & HyperCoffer~\cite{xia2013architecture}  & STC & STC & STC & - & - & C && \yes & \no & \yes \\
            & H-SVM~\cite{jin2011architectural,jin2015h} & STL & STL & STL & - & STL & - && \no & \yes & \yes \\
            & EqualVisor~\cite{deng2014equalvisor} & SL & SL & STL & - & STL & - && \no & \yes & \no \\
            & xu-cc15~\cite{xu2015architectural} & STC & STC & STC & ? & STC & C && \no & \no & \yes \\
            & wen-cf13~\cite{wen2013multi}       & STL & STL & STL & - & STL & C && \yes & \no & \yes \\
            & Komodo~\cite{ferraiuolo2017komodo} & SL & SL & STL & - & STL & - && \yes & \yes & \no \\
            & SANCTUARY~\cite{brasser2019sanctuary} & STL & STL & STL & - & STL & - && \yes & \no & \no \\
            & TrustICE~\cite{sun2015trustice}    & STL & STL & STL & - & STL  & - && \yes & \no & \no \\
            & HA-VMSI~\cite{zhu2017ha}           & SL & SL & STL & - & STL  & - && \yes & \yes & \yes \\
            & Sanctum~\cite{costan2016sanctum}   & STL & STL & STL & - & STL  & - && \yes & \no & \yes \\
            & TIMBER-V~\cite{weiser2019timber}   & STL & STL & STL & - & STL  & - && \no & \no & \yes \\
            & Keystone~\cite{lee2020keystone}    & STL & STL & STL & - & STL  & - && \yes & \no & \no \\
            & Penglai~\cite{feng2021scalable}    & STL & STL & STL & - & STL  & - && \no & \yes & \yes \\
            & CURE~\cite{bahami2021cure}         & STL & STL & STL & - & STL  & - && \yes & \no & \no \\
            & Iso-X~\cite{evtyushkin2014iso}     & STL & STL & STL & - & STL & - && \no & \yes & \yes \\
            & HyperWall~\cite{szefer2012architectural} & STL & STL & STL & - & STL & - && \no & \yes & \yes \\
            & Sancus~\cite{noorman2013sancus,noorman2017sancus} & STL & STL & STL & - & SL  & - && \yes & \no & \no \\
            & TrustLite~\cite{koeberl2014trustlite} & STL & STL & STL & - & -  & - && \yes & \no & \no \\
            & TyTan~\cite{brasser2015tytan}      & STL & STL & STL & - & -  & - && \yes & \no & \no \\
            & XOM~\cite{lie2000xom}              & STC & STC & STC & - & -  & C$^\dagger$ && \no & \no & \yes  \\
            & AEGIS~\cite{suh2003aegis}          & STC & STC & STC & - & STC  & C && \no & \no & \yes \\

        \bottomrule
    \end{tabular}
    \begin{tablenotes}
        \item [$\dagger$] No protection against replay attacks.
        \item [$\ddagger$] Memory encryption is optional but available for sale from the manufacturer. 
    \end{tablenotes}
    \caption{The various main memory isolation strategies used against individual adversaries. Note that some TEE designs use multiple strategies against different adversaries. Isolation strategies: cryptographic enforcement ($C$), logical enforcement ($L$), temporal partitioning ($T$), spatial partitioning ($S$), spatio-temporal partitioning ($ST$) , not considered (-). The required access control information that needs to be maintained by the TCB of the TEEs is split into three groups: memory protection unit (MPU), page table based, and extra metadata.}
    \label{tab:appendix:memoryisol}
    \end{threeparttable}
\end{table*}

\subsection{Verifiable Launch}

\subsubsection{Measurement}
\label{sec:appendix:measurement}
The root of trust for measurement roots a chain of trust. Every component in the chain is typically measured (and optionally verified) before it starts executing.
The measurement process for all these components in the chain of trust and the \enclave{} itself is similar. It entails mapping the binary executable of the component being measured to memory and computing one or more cryptographic hashes on that memory. Often, such measurements are built incrementally in compact form as a chain of cryptographic hashes on a page-by-page basis~\cite{mckeen2013innovative,evtyushkin2014iso,zhu2017ha}. %
Once the measurement of a component in the chain of trust is complete, its measurement must be stored securely such that an adversary cannot corrupt it. Some TEEs rely on special platform components such as TPM registers to store measurements~\cite{mccune2008flicker,mccune2008low,hunt2021power}. However, since TPMs are off-chip components that may be subject to attacks (e.g., by \advBus), some solutions store these measurements in on-chip registers~\cite{anati2013innovative,szefer2012architectural}. Furthermore, since TPMs  and on-chip solutions can only store a limited number of measurements, more recent works store measurements in memory~\cite{evtyushkin2014iso, anati2013innovative,kaplan2020amd,lebedev2018sanctumattest,tdxbasespec} within the \enclave{'s} TCB. %

\subsubsection{Attestation}
In general, the basic attestation process is well-understood. First, the \enclave{'s} underlying TCB constructs an \emph{attestation report} containing the \enclave{}'s measurement and its TCB. Then, the attestation report is integrity protected either through a cryptographic signature or a Message Authentication Code (MAC). Digital signatures  typically need additional infrastructure. For example, Intel SGX\cite{epidIntel2016} and Intel TDX\cite{tdxwhitepaper} require a separate Intel service called Intel Attestation Service (IAS) to verify a report containing the measurements. AMD SEV-SNP\cite{amdsevapi} signs all device-specific keys with an AMD root key and relies on a generic public key infrastructure. TPM-based attestation schemes also require third-party infrastructure such as a trusted third party to generate attestation keys~\cite{tpm}. In contrast, in local attestation, the underlying TCB typically creates a MAC-based report\cite{epidIntel2016} with a local key. We note that certain attestation proposals\cite{noorman2017sancus,brasser2015tytan,weiser2019timber} rely on using a MAC-based scheme even for remote attestation; Only SANCUS\cite{noorman2017sancus,muhlberg2015lightweight} discusses how the verifier obtains the symmetric key for verification. 

Historically, the only values included in an attestation report were the actual measurements of the \enclave{} and the TCB \cite{mccune2008flicker,mccune2008low} as well a nonce for freshness. However, more modern TEEs use extended attestation reports to include run-time attributes. For example, Intel SGX\cite{intelias} and AMD SEV-SNP\cite{kaplan2020amd} include a flag in the attestation report to indicate whether simultaneous multithreading (SMT) is enabled. Other types of information that could be added to an attestation report include software version numbers\cite{epidIntel2016,amdsevapi,tdxbasespec}, migration policies\cite{amdsevapi}, and TCB version when applicable. Finally, some TEE designs allow the \enclave{} to append some custom data (e.g., a public key certificate) into an attestation report which can be used later to establish a secure channel.

Most TEE designs include a basic set of primitives to support attestation, regardless of whether it is local or remote and independent of the types of information included in the report itself. First, TEE solutions must be able to securely generate and store attestation keys and protect them against unauthorized access/modification. The exact support required for the generation and storage of attestation keys varies based on the key hierarchy a given TEE solution adopts. Most signature-based attestation solutions include a \emph{platform identity key} that is used to establish the authenticity of the platform and one or more \emph{attestation keys} that are signed using the identity key. While attestation keys are usually generated on demand, the platform identity key is either generated once and stored permanently (e.g., Endorsement key in TPM based protocols~\cite{tpm}) or derived on every boot (e.g., based on Root Provisioning key in Intel SGX~\cite{epidIntel2016}). Furthermore, a hierarchy of attestation keys could be used to reflect updates to the mutable part(s) of the TCB in the attestation~\cite{epidIntel2016,lebedev2018sanctumattest,amdsevapi,tdxbasespec}.
Second, the TCB in TEE architectures may expose interfaces for requesting (and verifying) an attestation report. These interfaces may be exposed as new instructions (e.g., \cite{mckeen2013innovative,evtyushkin2014iso, tdxbasespec}) or services (e.g., \cite{mckeen2013innovative,costan2016sanctum,brasser2015tytan,koeberl2014trustlite}) when implemented in hardware or software respectively. 

\subsection{Memory Isolation}

\noindent \textbf{Memory Protection Unit (MPU)}
\label{sec:appendix:mpu}
While many different variants of memory protection units (MPU) exist, they all perform very similar tasks: they check the physical address and the access type (e.g., write or read) against access control information. Most MPU implementations can protect a limited number of memory regions at any point in time and are suitable for coarse-grained memory protections. 
In order to enable logical isolation, the access control information for the MPU must be configured and controlled solely by the \enclave{'s} TCB. 
Many modern academic TEEs rely on such an MPU to provide isolation\cite{lee2020keystone,bahami2021cure,weiser2019timber,brasser2019sanctuary,costan2016sanctum}.

\noindent \textbf{Memory Management Unit (MMU)}
In contrast to the coarse-grained MPU, MMUs allow for much finer-grained access control checks. An MMU is a logic block that converts a virtual address to a physical address using data structures called page tables. Page tables not only store these mappings but also additional security-sensitive information such as permissions (e.g., read, write, execute) alongside every entry. In TEE designs, MMUs can be used for logical memory isolation and offer better scalability and more fine-grained protections compared to MPU based approaches. However, the page tables themselves must be protected from the adversary.
There are a variety of ways to manage the page tables:
The most straightforward way to manage page tables is by letting the TCB control all the page tables not only for all the \enclave{s} but also all for the other untrusted software components (e.g., \advSSW, \advApp)~\cite{zhu2017ha,deng2014equalvisor,xia2013architecture,zhu2017ha}. In other proposals, the TCB controls only the page tables of all \enclave{s}; so, the TCB can ensure that the \enclave{} has access only to its own memory and prevent it from accessing the memory of other \enclave{s}. However, if the TCB does not control the page tables of \advSSW{}, it cannot prevent such a privileged attacker from accessing an \enclave{'s} memory. Examples of architectures that do this include Intel TDX~\cite{tdxbasespec}, ARM TZ\cite{armtz}, and ARM Realms\cite{armrealms}. In such designs, the MMU must use a set of \emph{secondary metadata} to enable access to \enclave{} memory based on the executing context (e.g., as recognized by the CPU mode or privilege level or other identifiers). This approach based on secondary metadata can also be used when the system design does not provide a way to set up trustworthy page tables for an \enclave{} itself~\cite{mckeen2013innovative, kaplan2020amd,evtyushkin2014iso,shinde2015podarch,costan2016sanctum}.

\noindent \textit{Translation Look-aside Buffers (TLBs):}
A translation look-aside buffer (TLB) essentially holds information about the MMU's recent virtual to physical translations. From a security perspective, it is essential to prevent the re-use of stale MMU translations by untrusted execution contexts. In modern processors, dedicated TLBs are usually available per thread of execution (e.g., per logical processor) and are hence, inherently spatially partitioned among concurrent threads. Furthermore, when a given logical processor is shared over time by different execution contexts, usually temporal isolation is used, i.e., TLBs are flushed on every context switch. However, since TLB  flushes can degrade performance, modern processors support the use of additional information about the execution context that owns each entry to enable partial TLB flushes during transitions\cite{riscv2019privspec,intelSDM,amdIsa,armv8}. This allows multiple execution contexts to share the TLB as a whole at any given point in time but prevents them from re-using each other's translations. Besides dedicated TLBs per logical processor and TLB flushes on context switches, some solutions~\cite{kaplan2020amd,tdxbasespec} rely on additional tags in their TLBs to protect translations belonging to \enclave{s}. Lastly, we did not see the usage of cryptographic isolation for TLBs, probably due to its potential cost and performance overheads. \moritz{TLB only need protection for MMU based approaches}

\noindent \textit{Cryptography Support in the Memory Controller:} 
While it is possible to include simple access control checks into the memory controller, the more common type of memory isolation enforcement at the memory controller is cryptographic. Encryption, integrity, and replay protection are are applied to data in the memory controller just before it exits the SoC. When used for protection against a \advBus, this isolation technique is typically implemented using a single pair of keys - one for encryption and the other for integrity protection (e.g., Intel SGX). Sometimes, critical ranges of memory are assigned separate keys as a defense-in-depth mechanism, but overall, defense against \advBus{} usually requires just a handful of keys (e.g., ARM CCA which uses a separate key per world). Metadata related to integrity and anti-replay protection are often stored in sequestered memory; the metadata corresponding to each memory transaction is retrieved/stored by memory subsystem usually in a way that is software-agnostic/entirely hardware managed (e.g., Intel SGX). The underlying micro-architectural components include a block that can generate these additional accesses to sequestered memory, a metadata cache for recently accessed memory, and filters to ensure only authorized access to this metadata.

When cryptographic enforcement is used to protect against software attacks (e.g., \advTEE, \advSSW) and malicious peripherals (\advPer), TEE architectures use a separate key per execution context. They commonly use a unique key per \enclave{} that is not available for use by other software adversaries or peripherals. Such solutions also require that information about the source of the memory request is carried along with every transaction all the way from the CPU to the memory controller. The memory controller uses this information to select the correct cryptographic key material for accessing the actual memory targeted by that transaction. This information about the execution context corresponding to the memory transaction is often retained in the TLBs and caches, and it is carried on all system fabrics to enforce access control within the SoC.

\end{document}